\documentclass[11pt]{article}
\usepackage{times}
\usepackage{amsmath,amsfonts}
\usepackage{epsfig,amssymb,amstext,amsthm}
\usepackage{graphicx,color} 
\usepackage{verbatim}
\usepackage{xspace,comment,calc}
\usepackage{listings,algpseudocode,algorithm}
\usepackage[margin=1cm]{caption}
\usepackage[shortlabels]{enumitem}

\newtheorem{theorem}{Theorem}[section]
\newtheorem{lemma}[theorem]{Lemma}
\newtheorem{claim}[theorem]{Claim}
{\theoremstyle{remark} }
{\theoremstyle{definition} \newtheorem{definition}[theorem]{Definition} }

\algtext*{EndWhile}
\algtext*{EndIf}
\algtext*{EndFor}

{\theoremstyle{remark} 
\newtheorem{myalg}{Algorithm}
\newtheorem*{kballalg}{Algorithm \boldmath $\kbalg(\D',\cL',k',\e)$}
\newtheorem*{primaldalg}{Algorithm \boldmath $\pdalg(\D',\cL',\z)$}
\newtheorem*{kballoalg}{Algorithm \boldmath $\kbo(\D',\cL',k',\e)$}
}
 
\newenvironment{proofof}[1]{\begin{proof}[Proof of #1]}{\end{proof}}

\newcommand{\al}{\ensuremath{\alpha}}
\newcommand{\dt}{\ensuremath{\delta}}
\newcommand{\sg}{\ensuremath{\sigma}}

\newcommand{\Om}{\ensuremath{\Omega}}
\newcommand{\e}{\ensuremath{\epsilon}}

\newcommand{\gm}{\ensuremath{\gamma}}

\newcommand{\kp}{\ensuremath{\kappa}}

\newcommand{\cA}{\ensuremath{\mathcal{A}}}

\newcommand{\cI}{\ensuremath{\mathcal I}}
\newcommand{\cS}{\ensuremath{\mathcal S}}
\newcommand{\cF}{\ensuremath{\mathcal F}}
\newcommand{\cD}{\ensuremath{\mathcal D}}
\newcommand{\cN}{\ensuremath{\mathcal N}}
\newcommand{\cL}{\ensuremath{\mathcal L}}
\newcommand{\cU}{\ensuremath{\mathcal U}}

\newcommand{\RRN}{\mathbb{R}_{\ge 0}}

\newcommand{\ceil}[1]{\ensuremath{\left\lceil#1\right\rceil}}

\newcommand{\sm}{\ensuremath{\setminus}}
\newcommand{\es}{\ensuremath{\emptyset}}
\newcommand{\sse}{\subseteq}

\newcommand{\OPT}{\mbox{\sc OPT}}

\newcommand{\iopt}{\ensuremath{O^*}}
\newcommand{\lpopt}{\ensuremath{\mathit{OPT}}}

\newcommand{\np}{{\em NP}\xspace}
\newcommand{\nphard}{\np-hard\xspace} 

\newcommand{\apx}{{\em APX}\xspace}

\newcommand{\z}{\ensuremath{z}}

\newcommand{\f}{\ensuremath{f}}

\newcommand{\cost}{\ensuremath{\mathit{cost}}}

\newcommand{\lbks}{\ensuremath{\mathsf{LB}k\mathsf{Sup}}\xspace}

\newcommand{\lbfl}{\ensuremath{\mathsf{LBFL}}\xspace}

\newcommand{\lbkco}{\ensuremath{\mathsf{LB}k\mathsf{CentO}}\xspace}
\newcommand{\lbkso}{\ensuremath{\mathsf{LB}k\mathsf{SupO}}\xspace}

\newcommand{\msr}{\ensuremath{k\mathsf{SR}}\xspace}
\newcommand{\msd}{\ensuremath{k\mathsf{SD}}\xspace}
\newcommand{\lbksr}{\ensuremath{\mathsf{LB}k\mathsf{SR}}\xspace}
\newcommand{\lbksro}{\ensuremath{\mathsf{LB}k\mathsf{SRO}}\xspace}
\newcommand{\kbs}{\ensuremath{k\text{-}\mathsf{BS}}\xspace}
\newcommand{\ksr}{\kbs}

\newcommand{\cc}{\ensuremath{\mathsf{CellC}}\xspace}

\newcommand{\rad}{\ensuremath{\mathsf{rad}}\xspace}

\newcommand{\uncov}{\ensuremath{\mathsf{uncov}}\xspace}
\newcommand{\uc}{\ensuremath{\mathsf{uc}}\xspace}
\newcommand{\outl}{\ensuremath{\mathit{Out}}\xspace}
\newcommand{\AT}{\ensuremath{\mathit{AT}}\xspace}
\newcommand{\AD}{\ensuremath{\mathit{AD}}\xspace}
\newcommand{\dist}{\ensuremath{\mathit{dist}}\xspace}

\newcommand{\pdalg}{\ensuremath{\mathsf{PDAlg}}\xspace}
\newcommand{\pdo}{\ensuremath{\mathsf{PDAlg^o}}\xspace}
\newcommand{\kbalg}{\ensuremath{k\text{-}\mathsf{BSAlg}}\xspace}
\newcommand{\kbo}{\ensuremath{k\text{-}\mathsf{BSAlg^o}}\xspace}

\newcommand{\A}{\ensuremath{\mathcal{A}}}
\newcommand{\B}{\ensuremath{\mathcal{B}}}
\newcommand{\I}{\ensuremath{\mathcal I}}
\newcommand{\F}{\ensuremath{\mathcal F}}
\newcommand{\D}{\ensuremath{\mathcal D}}

\newcommand{\Pc}{\ensuremath{\mathcal P}}
\newcommand{\Qc}{\ensuremath{\mathcal Q}}

\newcommand{\Sc}{\ensuremath{\mathcal S}}

\newcommand{\tq}{\ensuremath{\tilde q}}
\newcommand{\tx}{\ensuremath{\tilde x}}

\newcommand{\cball}{\ensuremath{B}}
\newcommand{\out}{\ensuremath{\mathsf{out}}}
\newcommand{\proj}{\ensuremath{\mu}}
\newcommand{\assign}{\ensuremath{\leftarrow}}
\newcommand{\nstar}{\ensuremath{\Sc}}

\title{Approximation Algorithms for Clustering Problems with Lower Bounds and Outliers%
\footnote{A preliminary version~\cite{AhmadianS16} appeared in the Proceedings of the
  International Colloquium on Automata, Languages and Programming (ICALP), 2016.}}

\author{
         Sara Ahmadian\thanks{{\tt \{sahmadian,cswamy\}@uwaterloo.ca}. 
         Dept. of Combinatorics and Optimization, Univ. Waterloo, Waterloo, ON N2L 3G1.
         Supported in part by NSERC grant 327620-09 and an NSERC Discovery Accelerator
         Supplement award.}   
\and
\addtocounter{footnote}{-1}
         Chaitanya Swamy\footnotemark
}

\date{}

\begin{document}

\maketitle

\begin{abstract}
We consider clustering problems with {\em non-uniform lower bounds and outliers}, and
obtain the {\em first approximation guarantees} for these problems.
We have a set $\F$ of facilities with lower bounds $\{L_i\}_{i\in\F}$ and a set $\D$ of
clients located in a common metric space $\{c(i,j)\}_{i,j\in\F\cup\D}$, and bounds $k$,
$m$. 
A feasible solution is a pair $\bigl(S\sse\cF,\sg:\D\mapsto S\cup\{\out\}\bigr)$, 
where $\sg$ specifies the client assignments,
such that $|S|\leq k$, $|\sg^{-1}(i)|\geq L_i$ for all $i\in S$, and 
$|\sg^{-1}(\out)|\leq m$.
In the {\em lower-bounded min-sum-of-radii with outliers} (\lbksro) problem, the objective
is to minimize $\sum_{i\in S}\max_{j\in\sg^{-1}(i)}c(i,j)$, and in the 
{\em lower-bounded $k$-supplier with outliers} (\lbkso) problem, the objective is to
minimize $\max_{i\in S}\max_{j\in\sg^{-1}(i)}c(i,j)$.

We obtain an approximation factor of $12.365$ for \lbksro, which improves to $3.83$ for
the non-outlier version (i.e., $m=0$). 
These also constitute the {\em first} approximation bounds 
for the min-sum-of-radii objective when we consider lower bounds and outliers 
{\em separately}. 
We apply the primal-dual method to the relaxation where we Lagrangify the $|S|\leq k$ 
constraint. The chief technical contribution and novelty of our
algorithm is that, departing from the standard paradigm used for such constrained
problems, we obtain an $O(1)$-approximation {\em despite the fact that we do not 
obtain a Lagrangian-multiplier-preserving algorithm for the Lagrangian relaxation}. We
believe that our ideas have 
{broader applicability to other clustering problems with outliers as well.}

We obtain approximation factors of $5$ and $3$ respectively for \lbkso and its non-outlier
version. These are the {\em first} approximation results for $k$-supplier with
{\em non-uniform} lower bounds. 
\end{abstract}

\section{Introduction}
Clustering is an ubiquitous problem that arises in many applications in different fields
such as data mining, machine learning, image processing, and bioinformatics.
Many of these problems involve finding a set $S$ of at most $k$ ``cluster centers'', and
an assignment $\sg$ mapping an underlying set $\D$ of data points located in some metric
space $\{c(i,j)\}$ to $S$, 
to minimize some objective function; 
examples include 
the {\em $k$-center} (minimize $\max_{j\in\D}c(\sg(j),j)$)~\cite{hochbaums85,hochbaums86}, 
{\em $k$-median} (minimize $\sum_{j\in\D}c(\sg(j),j)$)%
~\cite{charikar1999constant,jain2001approximation,li2013approximating,byrka2015improved}, and 
{\em min-sum-of-radii} (minimize $\sum_{i\in S}\max_{j:\sg(j)=i}c(i,j)$)%
~\cite{doddi2000approximation,charikar2001clustering} problems. 
Viewed from this perspective, clustering problems can often be 
viewed as {\em facility-location} problems, 
wherein an underlying set of clients 
that require service need to be assigned to facilities 
that provide service in a cost-effective fashion.
Both clustering and facility-location problems have been extensively studied 
in the Computer Science and Operations Research literature; see,
e.g.,~\cite{mirchandanif90,shmoys} in addition to the above
references. 

We consider clustering problems with (non-uniform) {\em lower-bound requirements} on the 
cluster sizes, and where a bounded number of points may be designated as {\em outliers}
and left unclustered.
One motivation for considering lower bounds comes from an {\em anonymity}
consideration. In order to achieve data privacy, \cite{Samarati00} proposed an
anonymization problem where we seek to perturb (in a specific way) some of (the attributes
of) the data points and then cluster them so that every cluster has at least $L$ identical
perturbed data points, thus making it difficult to identify the original data from the
clustering. 
As noted in~\cite{AggarwalFKMPZ05,aggarwal2010achieving}, 
this anonymization problem can be abstracted as a lower-bounded clustering problem 
where the clustering objective captures the cost of perturbing data.
Another motivation comes from a facility-location perspective, where (as in the case of
{\em lower-bounded facility location}), the lower bounds model that it is
infeasible or unprofitable to use services unless they satisfy a certain minimum demand
(see, e.g.,~\cite{LimWX06}).
Allowing outliers enables one to handle a common woe in clustering problems, 
namely that 
data points that are quite dissimilar from any other data
point can often disproportionately (and undesirably) degrade the quality of {\em any}
clustering of the {\em entire} data set; instead, the outlier-version allows one to
designate such data points as outliers and focus on the data points of interest.

Formally, adopting the facility-location terminology, our setup is as follows. 
We have a set $\F$ of facilities with lower bounds $\{L_i\}_{i\in\F}$ and a set $\D$ of
clients located in a common metric space $\{c(i,j)\}_{i,j\in\F\cup\D}$, and bounds $k$,
$m$. 
A feasible solution chooses a set $S\sse\F$ of at most $k$ facilities, and assigns each
client $j$ to a facility $\sg(j)\in S$, or designates $j$ as an outlier by setting
$\sg(j)=\out$ so that $|\sg^{-1}(i)|\geq L_i$ for all $i\in S$, and 
$|\sg^{-1}(\out)|\leq m$.
We consider two clustering objectives: 
minimize $\sum_{i\in S}\max_{j:\sg(j)=i}c(i,j)$, which yields the
{\em lower-bounded min-sum-of-radii with outliers} (\lbksro) problem, and 
minimize $\max_{i\in S}\max_{j:\sg(j)=i}c(i,j)$, which yields the 
{\em lower-bounded $k$-supplier with outliers} (\lbkso) problem. ($k$-supplier
denotes the facility-location version of $k$-center; the latter typically has $\F=\D$.)  
We refer to the non-outlier versions of the above problems (i.e., where $m=0$) as $\lbksr$ 
and $\lbks$ respectively.

\vspace*{-1ex}
\paragraph{Our contributions.}
We obtain the {\em first} results for clustering problems with {\em non-uniform lower
bounds and outliers}. We develop various techniques for tackling these problems  
using which we obtain {\em constant-factor approximation guarantees} for \lbksro and
\lbkso.   
Note that we need to ensure that none of 
the three types of {\em hard} constraints involved here---at
most $k$ clusters, non-uniform lower bounds, and at most $m$ outliers---are
violated, which is somewhat challenging.

We obtain an approximation factor of $12.365$ for \lbksro (Theorem~\ref{outthm},
Section~\ref{outl}), which improves to $3.83$ for the non-outlier version \lbksr
(Theorem~\ref{improvthm}, Section~\ref{nonoutl}). 
These also constitute the {\em first} approximation results 
for the min-sum-of-radii objective when we consider:
(a) lower bounds (even uniform bounds) but no outliers (\lbksr);
and (b) outliers but no lower bounds. 
Previously, an $O(1)$-approximation was known 
only in the setting where there are {\em no lower bounds and no outliers}
(i.e., $L_i=0$ for all $i$, $m=0$)~\cite{charikar2001clustering}.   

For the $k$-supplier objective (Section~\ref{lbksup}), we obtain an approximation factor
of $5$ for \lbkso (Theorem~\ref{lbksothm}), and $3$ for \lbks
(Theorem~\ref{lbksthm}). These are the {\em first} approximation results for the
$k$-supplier problem with non-uniform lower bounds. Previously, 
\cite{aggarwal2010achieving} obtained approximation factors of $4$ and $2$ respectively 
for \lbkso and \lbks for the special case of {\em uniform} lower bounds 
and when $\F=\D$.
Complementing our approximation bounds, we prove a factor-$3$ hardness of approximation
for \lbks (Theorem~\ref{lbkshard}), which shows that our approximation factor of
$3$ is optimal for \lbks. We also show (Appendix~\ref{append-equiv}) that \lbkso is
equivalent to the $k$-center version of the problem (where $\F=\D$). 
 
\vspace*{-1ex}
\paragraph{Our techniques.}
Our main technical contribution is an $O(1)$-approximation algorithm for \lbksro
(Section~\ref{outl}). 
Whereas for the non-outlier version \lbksr (Section~\ref{nonoutl}), one can follow an
approach similar to that of Charikar and Panigrahi~\cite{charikar2001clustering} for the
min-sum-of-radii problem without lower bounds or outliers, the presence of outliers
creates substantial difficulties whose resolution requires various novel ingredients.
As in~\cite{charikar2001clustering}, we view \lbksro as a {\em $k$-ball-selection} (\kbs)
problem of picking $k$ suitable balls (see Section~\ref{sec:lbksro}) 
and consider its LP-relaxation \eqref{primal2}. Let $\lpopt$ denote its optimal value.
Following the Jain-Vazirani (JV) template for $k$-median~\cite{jain2001approximation}, we
move to the version where we may pick any number of balls but incur a fixed cost of $\z$
for each ball we pick. The dual LP \eqref{dual2} has $\al_j$ dual variables for the
clients, which ``pay'' for $(i,r)$ pairs (where $(i,r)$ denotes the ball 
$\{j\in\D: c(i,j)\leq r\}$). 
For \lbksr (where $m=0$), as observed in~\cite{charikar2001clustering}, it is easy to
adapt the JV primal-dual algorithm for facility location to handle this fixed-cost version
of \kbs: we raise the $\al_j$s of uncovered clients until all clients are covered by some
fully-paid $(i,r)$ pair (see $\pdalg$). This yields a so-called 
{\em Lagrangian-multiplier-preserving} (LMP) $3$-approximation algorithm: if $F$ is the
primal solution constructed, then $3\sum_j\al_j$ can pay for $\cost(F)+3|F|\z$; 
hence, by varying $\z$, one can find two solutions $F_1$, $F_2$ for
nearby values of $\z$, and combine them to extract a low-cost \kbs-solution. 

The presence of outliers in \lbksro significantly complicates things. The natural
adaptation of the primal-dual algorithm is to now stop when at least $|\D|-m$ clients are 
covered by fully-paid $(i,r)$ pairs. But now, the dual objective involves a 
$-m\cdot\gm$ term, where $\gm=\max_j\al_j$, which potentially cancels the dual
contribution of (some) clients that pay for the last fully-paid $(i,r)$ pair, say
$\f$. Consequently, we {\em do not obtain an LMP-approximation}: if $F$ is the primal
solution we construct, we can only say that (loosely speaking) $3(\sum_j\al_j-m\cdot\gm)$
pays for $\cost(F\sm\f)+3|F\sm\f|\z$ (see Theorem~\ref{thm:outsumr} (ii)). 
In particular, this means that {\em even if the primal-dual algorithm returns a solution
with $k$ pairs, its cost need not be bounded}, an artifact that never arises in \lbksr (or
$k$-median). This in turn means that by combining the two solutions $F_1,F_2$ found for
$\z_1,\z_2\approx\z_1$, we only obtain a solution of cost $O(\lpopt+\z_1)$ 
(see Theorem~\ref{combA}).

Dealing with the case where $\z_1=\Om(\lpopt)$ 
is technically the most involved portion of our algorithm (Section~\ref{broutine}). 
We argue that in this case
the solutions $F_1$, $F_2$ (may be assumed to) have a very specific structure:
$|F_1|=k+1$, and every $F_2$-ball intersects at most one $F_1$-ball, and vice versa. We
utilize this structure to show that either we can find a good solution in a suitable
neighborhood of $F_1$ and $F_2$, or $F_2$ itself must be a good solution.

We remark that the above difficulties (i.e., the inability to pay for the last ``facility''
and the ensuing complications) also arise in the {\em $k$-median problem with outliers}.
We believe that our ideas also have implications for this problem
and should yield a much-improved approximation ratio for this problem. 
{(The current approximation ratio is a large (unspecified) constant~\cite{chen2008constant}.)}

For the $k$-supplier problem, \lbkso, we leverage the notion of skeletons and 
pre-skeletons defined by~\cite{cygan2014constant} in the context of 
{\em capacitated $k$-supplier with outliers}, wherein facilities have capacities instead
of lower bounds limiting the number of clients that can be assigned to them. Roughly
speaking, a skeleton $F\sse\F$ ensures there is a low-cost solution $(F,\sg)$. 
A pre-skeleton satisfies some of the properties of a skeleton. We show that if $F$ is a
pre-skeleton, then either $F$ is a skeleton or $F\cup\{i\}$ is a pre-skeleton for some
facility $i$. This allows one to find a sequence of facility-sets such that at least one
of them is a skeleton. For a given set $F$, one can check if $F$ admits a low-cost
assignment $\sg$, so this yields an $O(1)$-approximation algorithm.

\vspace*{-1ex}
\paragraph{Related work.}
There is a wealth of literature on clustering and facility-location (FL) problems (see,
e.g.,~\cite{mirchandanif90,shmoys}); we limit ourselves to the work that is relevant to
\lbksro and \lbkso. 

The only prior work 
on clustering problems to incorporate both lower bounds {\em and} 
outliers is by Aggarwal et al.~\cite{aggarwal2010achieving}. They obtain approximation
ratios of $4$ and $2$ respectively for
\lbkso and \lbks with {\em uniform} lower bounds and when $\F=\D$, which they consider as
a means of achieving anonymity. 
They also consider an alternate {\em cellular clustering} (\cc) objective and
devise an $O(1)$-approximation algorithm for lower-bounded \cc, 
again with uniform lower bounds, and mention that this can be extended to an
{$O(1)$-approximation for lower-bounded \cc with outliers.}

More work has been directed towards clustering problems that involve {\em either} outliers
or lower bounds (but not both), and here, clustering with outliers has received more
attention than lower-bounded clustering problems. 
Charikar et al.~\cite{CharikarKMN01} consider (among other problems) the
outlier-versions of the uncapacitated FL, $k$-supplier and $k$-median
problems. They devise constant-factor approximations for the first two problems, and a
bicriteria approximation for the $k$-median problem with outliers. They also proved a
factor-3 approximation hardness result for $k$-supplier with outliers. This nicely
complements our factor-3 hardness result for $k$-supplier with lower bounds but no
outliers. 
Chen~\cite{chen2008constant} obtained the first
true approximation for $k$-median with outliers via a sophisticated combination of the
primal-dual algorithm for $k$-median and local search that yields a large (unspecified)
$O(1)$-approximation. 
As remarked earlier, the difficulties that we overcome in designing our
$12.365$-approximation for \lbksro are similar in spirit to the difficulties that arise in
$k$-median with outliers, and we believe that our techniques should also help and
significantly improve the approximation ratio for this problem.
Cygan and Kociumaka~\cite{cygan2014constant} consider the 
{\em capacitated $k$-supplier with outliers} problem, and devise a $25$-approximation
algorithm. We leverage some of their ideas in developing our algorithm for \lbkso.

Lower-bounded clustering and FL problems remain largely unexplored and are not well 
understood. Besides \lbks (which has also been studied in Euclidean
spaces~\cite{ene2013fast}), 
another such FL problem that has been studied is 
{\em lower-bounded facility location}
(\lbfl)~\cite{karger2000building,guha2000hierarchical}, wherein we seek to open (any
number of) facilities (which have lower bounds) and assign each client $j$ to an open
facility $\sg(j)$ so as to minimize $\sum_{j\in\D}c(\sg(j),j)$. 
Svitkina~\cite{svitkina2010lower} obtained the first true approximation for
\lbfl, achieving an $O(1)$-approximation; the $O(1)$-factor was subsequently improved
by~\cite{ahmadian2012improved}. Both results apply to \lbfl with uniform lower bounds, and 
can be adapted to yield $O(1)$-approximations to the $k$-median variant (where we may open
at most $k$ facilities).  

We now discuss work related to our clustering objectives,
albeit that does not consider lower bounds or outliers.
Doddi et al.~\cite{doddi2000approximation} introduced the $k$-clustering
min-sum-of-diameters (\msd) problem, which is closely related to the $k$-clustering
min-sum-of-radii (\msr) problem: the \msd-cost is at least the \msr-cost, and at most 
twice the \msr-cost.
The former problem is somewhat better understood than the latter one.
Whereas the \msd problem is \apx-hard even for shortest-path metrics of unweighted graphs
(it is \nphard to obtain a better than 2 approximation~\cite{doddi2000approximation}), 
the \msr problem is only known to be \nphard for general metrics, and its complexity for
shortest-path metrics of unweighted graphs is not yet settled with only a quasipolytime
(exact) algorithm known~\cite{gibsonk10}.   
On the positive side, Charikar and Panigrahi~\cite{charikar2001clustering} devised the
first (and current-best) $O(1)$-approximation algorithms for these problems, obtaining
approximation ratios of $3.504$ and $7.008$ for the \msr and \msd problems respectively,  
and Gibson et al.~\cite{gibsonk10} obtain a quasi-PTAS for the \msr problem when
$\F=\D$. 
Various other results 
are known for specific metric spaces and when $\F=\D$, such as 
Euclidean spaces~\cite{gibsonk12,capoyleasr91} and metrics with bounded aspect
ratios~\cite{gibsonk10,behsazs15}. 

The $k$-supplier and $k$-center (i.e., $k$-supplier with $\F=\D$) objectives have a rich
history of study. Hochbaum and Shmoys~\cite{hochbaums85,hochbaums86} obtained
optimal approximation ratios of $3$ and $2$ for these problems respectively. 
Capacitated versions of $k$-center and $k$-supplier have also been studied: 
\cite{khuller2000capacitated} devised a 6-approximation for uniform capacities,
\cite{cygan2012lp} obtained the first $O(1)$-approximation for non-uniform capacities, and
this $O(1)$-factor was improved to 9 in~\cite{an2015centrality}.

Finally, our algorithm for \lbksro leverages the template based on Lagrangian relaxation
and the primal-dual method to emerge from the work
of~\cite{jain2001approximation,charikar2005improved} for the $k$-median problem.

\section{Minimizing sum of radii with lower bounds and outliers} \label{sec:lbksro}

Recall that in the {\em lower-bounded min-sum-of-radii with outliers} (\lbksro)
problem, we have a facility-set $\F$ and client-set $\D$ located in a metric space
$\{c(i,j)\}_{i,j\in\F\cup\D}$, 
lower bounds $\{L_i\}_{i\in\F}$, and 
bounds $k$ and $m$. A feasible solution is a pair $\bigl(S\sse\cF,\sg:\D\mapsto S\cup\{\out\}\bigr)$, where $\sg(j)\in S$ indicates that $j$ is assigned to facility $\sg(j)$, and $\sg(j)=\out$
designates $j$ as an outlier, such that $|\sigma^{-1}(i)| \geq L_i$ for all $i\in S$, and $|\sigma^{-1}(\out)|\leq m$.
The cost of such a solution is $\cost(S,\sigma) := \sum_{i\in S} r_i$, where
$r_i:=\max_{j\in\sigma^{-1}(i)} c(i,j)$ denotes the {\em radius} of facility $i$; the goal
is to find a solution of minimum cost. We use \lbksr to denote the non-outlier version where $m=0$.

It will be convenient to consider a relaxation of \lbksro that we call the 
{\em $k$-ball-selection} (\kbs) problem, which focuses on selecting at most $k$ balls 
centered at facilities of minimum total radius. More precisely, let 
$\cball(i,r):=\{j\in\cD: c(i,j) \leq r\}$ denote the ball of clients 
centered at $i$ with radius $r$. Let $c_{\max}=\max_{i\in\F,j\in\D}c(i,j)$. 
Let $\cL_i := \{(i,r): |\cball(i,r)| \geq L_i\}$, and $\cL := \bigcup_{i\in \cF}
\cL_i$. The goal in \kbs is to find a set $F \sse \cL$ with $|F|\leq k$ and 
$\bigl|\cD\sm\bigcup_{(i,r)\in F} \cball(i,r)\bigr|\leq m$ so that 
$\cost(F) := \sum_{(i,r) \in F} r$ is minimized. (When formulating the LP-relaxation of the \kbs-problem,
we equivalently view $\cL$ as containing 
only pairs of the form $(i,c(i,j))$ for some client $j$, 
which makes $\cL$ finite.) 
It is easy to see that any \lbksro-solution yields a \kbs-solution of no greater
cost. The key advantage of working with \kbs is that 
we do not explicitly consider the lower bounds (they are folded into the $\cL_i$s) and 
we do not require the balls $\cball(i,r)$ for $(i,r)\in F$ to be disjoint. 
While a \kbs-solution $F$ need not directly translate to a feasible \lbksro-solution, one
can show that it does yield a feasible \lbksro-solution of cost at most
$2\cdot\cost(F)$. We prove a stronger version of this statement in
Lemma~\ref{dcost}. In the following two sections, 
we utilize this relaxation to devise the 
{\em first} constant-factor approximation algorithms for for \lbksr and \lbksro. 
To our knowledge, our algorithm is also the first $O(1)$-approximation algorithm for the  
outlier version of the min-sum-of-radii problem {\em without} lower bounds.

We consider an LP-relaxation for the \kbs-problem, and to round a fractional
\kbs-solution to a good integral solution, we need to preclude radii that are much larger
than those used by an (integral) optimal solution. 
We therefore ``guess'' the 
$t$ facilities in the optimal solution with the largest radii, and their radii, where  
$t\geq 1$ is some constant. That is, we enumerate over all
$O\bigl((|\F|+|\D|)^{2t}\bigr)$ choices 
$F^O=\{(i_1,r_1),\ldots,(i_t,r_t)\}$ of $t$ $(i,r)$ pairs from $\cL$. 
For each such selection, we 
set $\D'=\D\sm\bigcup_{(i,r)\in F^O}\cball(i,r)$, 
$\cL'=\{(i,r)\in\cL: r\leq\min_{(i,r)\in F^O}r\}$ and $k'=k-|F^O|$, and run
our \kbs-algorithm on the modified \kbs-instance $(\F,\D'.\cL',c,k',m)$ to obtain
a \kbs-solution $F$. 
We translate $F\cup F^O$ to an \lbksro-solution, and return the best of these solutions. 
The following lemma, and the procedure described therein, is repeatedly used to
bound the cost of translating $F\cup F^O$ to a feasible \lbksro-solution. 
We call pairs $(i,r), (i',r')\in \cF\times\RRN$ {\em non-intersecting}, if $c(i,i')>r+r'$,
and {\em intersecting} otherwise. Note that $\cball(i,r)\cap \cball(i',r')=\emptyset$ if
$(i,r)$ and $(i',r')$ are non-intersecting. For a set $P\sse\cF\times\RRN$ of pairs,
define $\proj(P):=\{i\in\cF: \exists r\text{ s.t. }(i,r)\in P\}$.

\begin{lemma}\label{lem:doublecost} \label{dcost}
Let $F^O\sse\cL$, and $\D',\cL',k'$ be as defined above.
Let $F\sse\cL$ be a \kbs-solution for the \kbs-instance $(\cF,\cD',\cL',c,k',m)$.
Suppose for each $i\in\proj(F)$, we have a radius $r'_i\leq\max_{r:(i,r)\in F}r$ such that
the pairs in $U:=\bigcup_{i\in\proj(F)}(i,r'_i)$ are non-intersecting and
$U\sse\cL$. Then there exists a feasible \lbksro-solution $(S,\sigma)$ with 
$\cost(S,\sigma)\leq\cost(F)+\sum_{(i,r)\in F^O}2r$. 
\end{lemma}

\begin{proof}
Pick a maximal subset $P\sse F^O$ to add to $U$ such that all pairs in $U'=U\cup P$ are
non-intersecting. For each $(i,r)\in F^O\sm P$, define $\kp(i,r)$ to be some 
intersecting pair $(i',r')\in U'$. 
Define $S=\proj(U')$. 
Assign each client $j$ to $\sg(j)\in S$ as follows. 
If $j\in \cball(i,r)$ for some $(i,r)\in U'$, set $\sigma(j)=i$. 
Note that $U'\sse \cL$, so this satisfies the lower bounds for all $i\in S$.
Otherwise, if $j\in \cball(i,r)$  for some $(i,r)\in F$, set $\sg(j)=i$. 
Otherwise, if $j\in \cball(i,r)$ for some $(i,r)\in F^O\sm P$ and $(i',r')=\kp(i,r)$, set
$\sg(j)=i'$. 
Any remaining unassigned client is not covered by the balls corresponding to pairs in
$F\cup F^O$. 
There are at most $m$ such clients, and we set $\sigma(j) = \out$ for each such client
$j$. 
Thus $(S,\sg)$ is a feasible \lbksro-solution.

For any $i\in S$ and $j\in\sg^{-1}(i)$ 
either $j\in B(i,r)$ for some $(i,r)\in F\cup U'$, or $j\in \cball(i',r')$ 
where $\kp(i',r')=(i,r)\in U'$, in which case $c(i,j)\leq r+2r'$.
{So $\cost(S,\sigma)\leq\cost(F) + \sum_{(i,r)\in F^O}2r$.}
\end{proof}

\subsection{Approximation algorithm for \lbksr} \label{nonoutl}
We now present our algorithm for the non-outlier version, \lbksr, which will introduce
many of the ideas 
underlying our algorithm for \lbksro described in Section~\ref{outl}. 
Let $\iopt$ denote the cost of an optimal solution to the given \lbksr instance. 

As discussed above, for each selection of $(i_1,r_1),\ldots,(i_t,r_t)$ of $t$ pairs, we do 
the following.
We set $\D'=\D\sm\bigcup_{p=1}^t\cball(i_p,r_p)$, 
$\cL'=\{(i,r)\in\cL: r\leq R^*:=\min_{p=1,\ldots,t}r_p\}$, $k'=k-t$, 
and consider the \kbs-problem of picking at most $k'$ pairs from $\cL'$
whose corresponding balls cover $\D'$ incurring minimum cost (but our algorithm $\kbalg$
will return pairs from $\cL$). We consider 
the following natural LP-relaxation \eqref{primal1} of this problem, and its dual
\eqref{dual1}.  

\vspace*{-4ex}

{\centering
\hspace*{-2ex}
\begin{minipage}[t]{0.45\textwidth}
\begin{alignat}{2}
\min & \quad & \sum_{(i,r) \in \cL'} r \cdot y_{i,r} & \tag{P$_1$} \label{primal1} \\
\text{s.t.} && \sum_{(i,r)\in\cL': j\in \cball(i,r)} y_{i,r} & \geq  1 
\quad \ \ \forall j \in \cD' \notag \\
&& \sum_{(i,r) \in \cL'} y_{i,r} & \le k' \label{kbnd} \\[-0.5ex]
&& y & \geq 0. \notag
\end{alignat} 
\end{minipage}
\begin{minipage}[t]{0.55\textwidth}
\begin{alignat}{2}
\max & \quad & \sum_{j\in \cD'} \alpha_j - k'\cdot z & \tag{D$_1$} \label{dual1} \\
\text{s.t.} && \sum_{j\in \cball(i,r)\cap\D'}\alpha_j - z & \leq  r 
\quad \ \ \forall (i,r)\in\cL' \label{eq:pairopen} \\
&& \al,z & \geq 0 \notag.
\end{alignat}
\end{minipage}
}

\smallskip 
\noindent
If \eqref{primal1} is infeasible then we discard this choice of $t$ pairs and move on to
the next selection. So we assume \eqref{primal1} is feasible in the remainder of this
section. 
Let $\lpopt$ denote the common optimal value of \eqref{primal1} and \eqref{dual1}.
As in the JV-algorithm for $k$-median,
we Lagrangify 
constraint \eqref{kbnd} and consider the unconstrained problem where we do not bound the
number of pairs we may pick, 
but we incur a fixed cost $\z$ for each pair $(i,r)$ that we pick (in addition to $r$). It 
is easy to adapt the JV primal-dual algorithm for facility
location~\cite{jain2001approximation} to devise a simple 
{\em Lagrangian-multiplier-preserving} (LMP) 3-approximation algorithm for this problem
(see $\pdalg$ and Theorem~\ref{thm:sumr}).
We use this LMP algorithm within a binary-search procedure for $\z$ to obtain two solutions
$F_1$ and $F_2$ with $|F_1|\leq k<|F_2|$, and show that these can be ``combined'' to
extract a \kbs-solution $F$ of cost at most $3.83\cdot\lpopt+O(R^*)$ (Lemma~\ref{improvcomb}). 
This combination step is more involved than in $k$-median. The main idea here is
to use the $F_2$ solution as a guide to merge some $F_1$-pairs. We cluster the $F_1$ pairs
around the $F_2$-pairs and setup a {\em covering-knapsack problem} whose solution
determines for each $F_2$-pair $(i,r)$, whether to ``merge'' the $F_1$-pairs clustered
around $(i,r)$ or select all these $F_1$-pairs (see step B2). 
Finally, we add back 
the pairs $(i_1,r_1),\ldots(i_t,r_t)$ selected earlier 
and apply Lemma~\ref{dcost} to obtain an \lbksr-solution. As required by
Lemma~\ref{dcost}, to aid in this translation,
our \kbs-algorithm returns, along with $F$, a suitable radius $\rad(i)$ for every facility 
$i\in\proj(F)$.  
This yields a $(3.83+\e)$-approximation algorithm (Theorem~\ref{improvthm}).

While our approach is similar to 
the one in~\cite{charikar2001clustering} for the 
min-sum-of-radii problem {\em without} lower bounds (although our combination step is 
notably simpler), 
an important distinction that arises is the following. 
In the absence of lower bounds, the ball-selection problem \kbs is {\em equivalent}
to the min-sum-of-radii problem, but (as noted earlier) this is no longer the
case when we have lower bounds since in \kbs we do not insist that the balls 
we pick be disjoint. 
Consequently, moving from overlapping balls in a \kbs-solution to an \lbksr-solution
incurs, in general, a factor-2 blowup in the cost (see Lemma~\ref{dcost}). 
It is interesting that we are able to avoid this blowup and obtain an
approximation factor that is quite close to the approximation factor (of $3.504$) achieved 
in~\cite{charikar2001clustering} for the min-sum-of-radii problem without lower bounds.  

We now describe our algorithm in detail and proceed to analyze it. We describe a slightly
simpler $(6.183+\e)$-approximation algorithm below (Theorem~\ref{nonoutthm}). We sketch the
ideas behind the improved approximation ratio at the end of this section and defer the
details to Appendix~\ref{append-improved}.

\begin{figure}[t!]
\begin{center}
\includegraphics[width=0.55\textwidth]{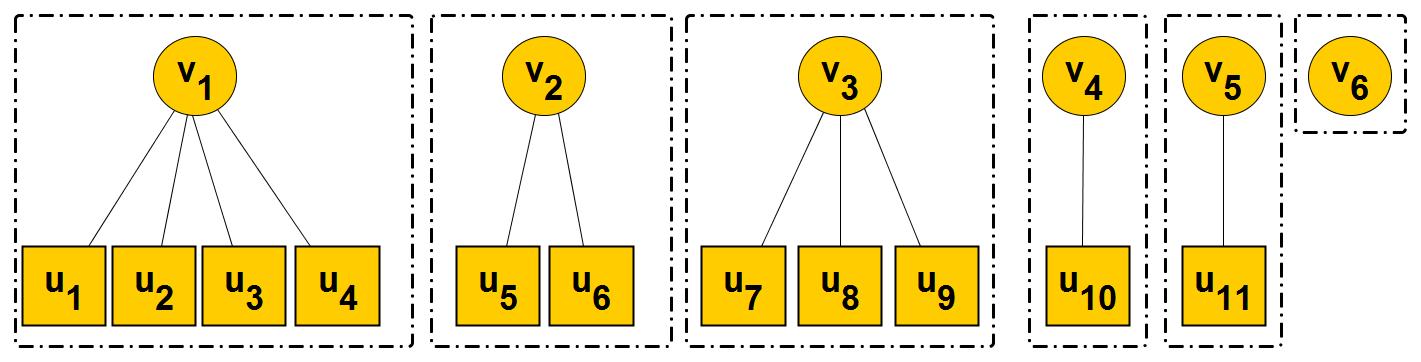}
\caption{An example of stars formed by $F_1$ and $F_2$ where $F_1 =
  \{u_1,u_2,\ldots,u_{11}\}$ and $F_2 = \{v_1,v_2,\ldots,v_6\}$ depicted by squares and circles, respectively.}
\label{fig:stars}
\end{center}
\vspace*{-4ex}
\end{figure}

{\small \vspace{5pt} \hrule
\begin{myalg} \label{nonoutalg}

Input: An \lbksr-instance $\I=\bigl(\F,\D,\{L_i\},\{c(i,j)\},k\bigr)$, parameter $\e>0$.

\noindent
Output: A feasible solution $(S,\sg)$.
\begin{enumerate}[label=A\arabic*., topsep=0.5ex, itemsep=0ex, labelwidth=\widthof{A3.},
    leftmargin=!] 
\item Let $t=\min\bigl\{k,\ceil{\frac{1}{\e}}\bigr\}$.
For each set $F^O\sse\cL$ 
with $|F^O|=t$, do the following. 
\begin{enumerate}[label*=\arabic*., topsep=0ex, itemsep=0ex, labelwidth=\widthof{A1.3.},
    leftmargin=!] 
\item Set $\D'=\D\sm\bigcup_{(i,r)\in F^O}\cball(i,r)$, 
  $\cL'=\{(i,r)\in\cL:r\leq R^*=\min_{(i,r)\in F^O}r\}$, and $k'=k-t$.
\item If \eqref{primal1} is infeasible, then reject this guess and move to the next set
  $F^O$. If $\D'\neq\es$, run $\kbalg(\D',\cL',k',\e)$ to obtain
  $\bigl(F,\{\rad(i)\}_{i\in F}\bigr)$; else set $(F,\rad)=(\es,\es)$. 
\item Apply the procedure in Lemma~\ref{dcost} taking $r'_i=\rad(i)$ for all
$i\in\proj(F)$ to obtain $(S,\sg)$.
\end{enumerate}

\item Among all the solutions $(S,\sg)$ found in step A2, return the one with smallest
  cost. 
\end{enumerate}
\newpage
\begin{kballalg}

Output: $F\sse\cL$ with $|F|\leq k'$, a radius $\rad(i)$ for all $i\in\proj(F)$.
\begin{enumerate}[label=B\arabic*., topsep=0.5ex, itemsep=0ex, labelwidth=\widthof{B3.},
    leftmargin=!] 
\item {\bf Binary search for \boldmath $\z$.\ }
\begin{enumerate}[label*=\arabic*., topsep=0ex, itemsep=0ex, labelwidth=\widthof{B1.2.},
    leftmargin=!] 
\item Set $\z_1 = 0$ and $z_2= 2k'c_{\max}$. 
For $p=1,2$, let $(F_p,\{\rad_p(i)\},\al^p)\assign\pdalg(\D',\cL',\z_p)$,
and let $k_p=|F_p|$.
If $k_1\leq k'$, stop and return $\bigl(F_1,\{\rad_1(i)\}\bigr)$.
We prove in Theorem~\ref{thm:sumr} that $k_2\leq k'$; if $k_2=k'$, stop and return
$\bigl(F_2,\{\rad_2(i)\}\bigr)$. 

\item Repeat the following until $z_2-z_1\leq \delta_z=\frac{\e\lpopt}{3n}$, 
where $n=|\cF|+|\cD|$. 
Set $z=\frac{z_1+z_2}{2}$. Let $(F,\{\rad(i)\},\al)\assign\pdalg(\D',\cL',\z)$. 
If $|F|=k'$, stop and return $\bigl(F,\{\rad(i)\}\bigr)$;
if $|F|>k'$, update $z_1\assign z$ and $(F_1,\rad_1,\al^1)\assign (F,\rad,\al)$,
else update $z_2\assign z$ and $(F_2,\rad_2,\al^2)\assign(F,\rad,\al)$. 
\end{enumerate}

\item {\bf Combining \boldmath $F_1$ and $F_2$.\ } 
Let $\pi:F_1\mapsto F_2$ be any map such that $(i',r')$ and $\pi(i',r')$ intersect 
$\forall (i',r')\in F_1$. (This exists since every $j\in\D'$ is covered by $\cball(i,r)$ for
some $(i,r)\in F_2$.) Define star $\nstar_{i,r}=\pi^{-1}(i,r)$ for all $(i,r)\in F_2$ (see
Fig.~\ref{fig:stars}).  
Solve the following {\em covering-knapsack LP}. 
\begin{alignat*}{2}
\min & \quad & \sum_{(i,r)\in F_2}\Bigl(x_{i,r}(2r+
{\textstyle \sum_{(i',r')\in\nstar_{i,r}}}2r') & +
(1-x_{i,r}){\textstyle \sum_{(i',r')\in\nstar_{i,r}}} r'\Bigr) \tag{C-P} \label{cklp} \\
\text{s.t.} & \quad
& \sum_{(i,r)\in F_2}\bigl(x_{i,r}+|\nstar_{i,r}|(1-x_{i,r})\bigr) & \leq k, 
\qquad 0 \leq x_{i,r} \leq 1 \quad \forall (i,r)\in F_2. 
\end{alignat*}
Let $x^*$ be an extreme-point optimal solution to \eqref{cklp}. The variable $x_{(i,r)}$
has the following interpretation. If $x^*_{i,r}=0$, then we select all pairs in
$\nstar_{i,r}$. Otherwise, if $\nstar_{i,r}\neq\es$, we pick a pair in
$(i',r')\in\nstar_{i,r}$, and include $(i',2r+r'+\max_{(i'',r'')\in\nstar_{i,r}\sm\{(i',r')\}}2r'')$ in 
our solution. Notice that by expanding the radius of $i'$ to
$2r+r'+\max_{(i'',r'')\in\nstar_{i,r}\sm\{(i',r')\}}2r''$, we cover all the clients in 
$\bigcup_{(i'',r'')\in\nstar_{i,r}}\cball(i'',r'')$. 
Let $F'$ be the resulting set of pairs. 

\item If $\cost(F_2)\leq\cost(F)$, return $(F_2,\rad_2)$, else
return $\bigl(F',\{\rad_1(i)\}_{i\in\proj(F')}\bigr)$. 
\end{enumerate}
\end{kballalg}

\begin{primaldalg} 

Output: $F\sse\cL$, radius $\rad(i)$ for all $i\in\proj(F)$, dual solution $\al$. 
\begin{enumerate}[label=P\arabic*., topsep=0ex, itemsep=0ex, labelwidth=\widthof{B3.},
    leftmargin=!] 
\item \textbf{Dual-ascent phase.\ } 
Start with $\al_j=0$ for all $j\in\D'$,
$\D'$ as the set of {\em active clients}, and the set $T$ of {\em tight pairs} initialized
to $\emptyset$. 
We repeat the following until all clients become inactive:
we raise the $\alpha_j$s of all active clients uniformly 
until constraint \eqref{eq:pairopen} becomes tight for some $(i,r)$;
we add $(i,r)$ to $T$ and mark all active clients in $\cball(i,r)$ as inactive.  

\item \textbf{Pruning phase.\ } 
Let $T_I$ be a maximal subset of non-intersecting pairs in $T$ picked by a greedy
algorithm that scans pairs in $T$ in non-increasing order of radius. Note that for each
$i\in\proj(T_I)$, there is exactly one pair $(i,r)\in T_I$. We set $\rad(i)=r$, 
and $r_i=\max\ \{c(i,j): j\in \cball(i',r'),\  (i',r')\in T,\ r'\leq r,\  
(i',r')\text{ intersects }(i,r)$ $((i',r')\text{ could be }(i,r))\}$. 
Let $F=\{(i,r_i)\}_{i\in\proj(T_I)}$. 
Return $F$, $\{\rad(i)\}_{i\in\proj(T_I)}$, and $\al$.
\end{enumerate}
\end{primaldalg}
\end{myalg}
\hrule
}

\paragraph{Analysis.} We prove the following result. 

\begin{theorem} \label{nonoutthm}
For any $\e>0$, Algorithm~\ref{nonoutalg} returns a feasible \lbksr-solution of cost at
most $\bigl(6.1821+O(\epsilon)\bigr)\iopt$ in time $n^{O(1/\epsilon)}$.
\end{theorem}

We first prove that $\pdalg$ is an LMP 3-approximation algorithm, i.e., its output
$(F,\al)$ satisfies $\cost(F)+3|F|\z\leq 3\sum_{j\in\D'}\al_j$
(Theorem~\ref{thm:sumr}). Utilizing this, we analyze $\kbalg$, in particular, the output 
of the combination step B2, and argue that $\kbalg$ returns a feasible solution of cost at
most $\bigl(6.183+O(\e)\bigr)\cdot\lpopt+O(R^*)$ (Theorem~\ref{combcost}). 
For the right choice of $F^O$, combining this with Lemma~\ref{dcost} yields
Theorem~\ref{nonoutthm}. 

\begin{theorem}\label{thm:sumr}
Suppose $\pdalg(\D',\cL',\z)$ returns $(F,\{\rad(i)\},\al)$. 
Then 
\begin{enumerate}[(i), itemsep=0ex, topsep=0.5ex]
\item the balls corresponding to $F$ cover $\D'$, \quad
\item $\cost(F)+3|F|z\leq 3\sum_{j\in \cD'}\alpha_j\leq 3(\lpopt+k'\z)$, 
\item $\bigl\{(i,\rad(i))\bigr\}_{i\in\proj(F)}\sse\cL'$, is a set of non-intersecting
pairs, and $\rad(i)\leq r_i\leq 3R^*\ \forall i\in\proj(F)$, 
\item if $|F|\geq k'$ then $\cost(F)\leq 3\cdot\lpopt$; if $|F|>k'$, then $\z\leq\lpopt$. 
{(Hence, $k_2\leq k'$ in step B1.1.)}
\end{enumerate}
\end{theorem}
 
\begin{proof}
We prove parts (i)---(iii) first.
Note that $\bigl\{(i,\rad(i))\bigr\}_{i\in\proj(F)}$ is $T_I$ (by definition).
Consider a client $j\in\D'$ and let $(i',r')$ denote the pair in $T$ that causes $j$ to
become inactive. Then there must be a pair $(i,r)\in T_I$ that intersects $(i',r')$
such that $r\geq r'$ (we could have $(i,r)=(i',r')$). Since by definition 
$r_i\geq c(i,j)$, $j\in B(i,r_i)$. Also, $c(i,i')\leq r+r'$.
So if $j$ is the client that determines $r_i$, then 
$r_i = c(i,j)\leq c(i',i)+c(i,j)\leq 2r'+r\leq 3r\leq 3R^*$.  

All pairs in $T_I$ are tight and non-intersecting. 
So for every $i\in\proj(F)$, there must be some $j\in\cball(i,\rad(i))\cap\D'$ with
$c(i,j)=\rad(i)$, so $\rad(i)\leq r_i$. 
Since $|F|=|T_I|$,
\begin{equation*} 
\cost(F)+3|F|\z=\sum_{(i,r)\in T_I}(r_i+3\z)\leq\sum_{(i,r) \in T_I} (3r+3\z) 
= \sum_{\substack{(i,r)\in T_I \\ j\in \cball(i,r)\cap\D'}}3\alpha_j \leq \sum_{j\in \cD'}3\alpha_j
\leq 3(\lpopt+k'\z). 
\end{equation*}
The last inequality above follows since $(\al,\z)$ is a feasible solution to \eqref{dual1}.

Rearranging the bound yields $3(|F|-k')\z\leq 3\cdot\lpopt-\cost(F)$, so 
when $|F|\geq k'$, we have $\cost(F)\leq 3\cdot\lpopt$, and when $|F|>k'$, we have
$\z\leq\lpopt$. 

Recall that in step B1.1, $k_2$ is the number of pairs returned by $\pdalg$ for
$\z=2k'c_{\max}$. So the last statement follows since $\lpopt\leq k'c_{\max}$, as all balls
in $\cL'$ have radius at most $c_{\max}$ and any feasible solution to \eqref{primal1}
satisfies $\sum_{(i,r)\in\cL'}y_{i,r}\leq k'$.
\end{proof}

Let $\bigl(F,\{\rad(i)\}\bigr)=\kbalg(\D',\cL',k')$. 
If $\kbalg$ terminates in step B1, 
then $\cost(F)\leq 3\cdot\lpopt$ due to part (ii) of Theorem~\ref{thm:sumr}, 
so assume otherwise. 
Let $a,b\geq 0$ be such that $ak_1+bk_2=k'$, $a+b=1$. Let $C_1=\cost(F_1)$ and
$C_2=\cost(F_2)$. 
Recall that $(F_1,\rad_1,\al^1)$ and $(F_2,\rad_2,\al^2)$ are the outputs of
$\pdalg$ for $\z_1$ and $\z_2$ respectively. 

\begin{claim} \label{fraccost}
We have $aC_1+bC_2\leq(3+\e)\lpopt$.
\end{claim}

\begin{proof}
By part (ii) of Theorem~\ref{thm:sumr}, we have $C_1+3k_1\z_1\leq 3(\lpopt+k'\z_1)$ and
$C_2+3k_2\z_2\leq 3(\lpopt+k'\z_2)$. Combining these, we obtain
$$
aC_1+bC_2\leq 3\lpopt+3k'(a\z_1+b\z_2)-3(ak_1\z_1+bk_2\z_2)
\leq 3(\lpopt+k'\z_2)-3k'\z_2+3ak_1\dt_z \leq (3+\e)\lpopt.
$$
The second inequality follows since $0\leq\z_2-z_1\leq\dt_z$.
\end{proof}

\begin{theorem} \label{combcost}
$\kbalg(\D',\cL',k')$ returns a feasible solution $\bigl(F,\{\rad(i)\}\bigr)$ with 
$\cost(F)\leq\bigl(6.183+O(\e)\bigr)\cdot\lpopt+O(R^*)$ 
where $\bigl\{(i,\rad(i))\}_{i\in\proj(F)}\sse\cL'$ is a set of non-intersecting pairs.
\end{theorem}

\begin{proof}
The radii $\{\rad(i)\}_{i\in\proj(F)}$ are 
simply the radii obtained from some execution of $\pdalg$, so \linebreak
$\bigl\{(i,\rad(i))\bigr\}_{i\in\proj(F)}\sse\cL'$ and comprises non-intersecting pairs.  
If $\kbalg$ terminates in step B1, we have argued a better bound on
$\cost(F)$. 
If not, and we return $F_2$, the cost incurred is $C_2$. 

Otherwise, we return the solution
$F'$ found in step B2. Since \eqref{cklp} has only one constraint 
in addition to the bound constraints $0\leq x_{i,r}\leq 1$, 
the extreme-point optimal solution $x^*$ has at most one fractional component, and if it
has a fractional component, then 
$\sum_{(i,r)\in F_2}\bigl(x^*_{i,r}+|\nstar_{i,r}|(1-x^*_{i,r})\bigr)=k'$.  
For any $(i,r)\in F_2$ with $x^*_{i,r}\in\{0,1\}$, the number of pairs we include is
exactly $x^*_{i,r}+|\nstar_{i,r}|(1-x^*_{i,r})$, and the total cost of these pairs is at
most the contribution to the objective function of \eqref{cklp} from the $x^*_{i,r}$ and
$(1-x^*_{i,r})$ terms. 
If $x^*$ has a fractional component $(i',r')\in F_2$, then 
$x^*_{i',r'}+|\nstar_{i',r'}|(1-x^*_{i',r'})$ is a {\em positive} integer. Since we
include at most one pair for $(i',r')$, this implies that $|F'|\leq k'$.
The cost of the pair we include is at most $15R^*$, since all 
$(i,r)\in F_1\cup F_2$ satisfy $r\leq 3R^*$. 
Therefore, $\cost(F')\leq\lpopt_{\text{\ref{cklp}}}+15R^*$. 
Also, $\lpopt_{\text{\ref{cklp}}}\leq 2bC_2+(2b+a)C_1=2bC_2+(1+b)C_1$, since setting
$x_{i,r}=b$ for all $(i,r)\in F_2$ yields a feasible solution to \eqref{cklp} of this
cost. 

So when we terminate in step B3, we return a solution $F$ with 
$\cost(F)\leq\min\{C_2,2bC_2+(1+b)C_1+15R^*\}$. 
The following claim (Claim~\ref{clm:costab}) shows that 
$\min\{C_2,2bC_2+(1+b)C_1\}\leq 2.0607(aC_1+bC_2)$ for all $a,b\geq 0$ with
$a+b=1$. 
Combining this with Claim~\ref{fraccost} yields the bound in the theorem. 
\end{proof}

\begin{claim} \label{clm:costab}
$\min\{C_2, 2bC_2+(1+b)C_1\} \leq (\frac{b+1}{3b^2-2b+1})(aC_1 +bC_2)\leq 2.0607(aC_1+bC_2)$
for all $a,b\geq 0$ such that $a+b=1$.
\end{claim}

\begin{proof}
Since the minimum is less than any convex combination,
\begin{eqnarray*}
\min(C_2, 2b C_2+b C_1 + C_1) &\leq &\frac{3b^2-b}{3b^2-2b+1}C_2 + \frac{1-b}{3b^2-2b+1}(2bC_2+bC_1+C_1) \\
 &= &\frac{(1-b)(1+b)}{3b^2-2b+1}C_1 + \frac{b^2+b}{3b^2-2b+1}C_2 = \frac{b+1}{3b^2-2b+1}((1-b)C_1+bC_2)
\end{eqnarray*}
Since $a = 1-b$, the first inequality in the claim follows.

The expression $\frac{b+1}{3b^2-2b+1}$ is maximized at $b=-1+\sqrt{2}$, and has value
$1+\frac{3}{2\sqrt{2}} \approx 2.0607$, which yields the second inequality in the claim.  
\end{proof}

Now we have all the ingredients needed for proving the main theorem of this section. 

\begin{proofof}{Theorem~\ref{nonoutthm}}
It suffices to show that when the selection $F^O=\{(i_1,r_1),\ldots(i_t,r_t)\}$ in step A1 
corresponds to the $t$ facilities in an optimal solution with largest radii, we obtain the
desired approximation bound. In this case, if $t=k$, then $F^O$ is an optimal
solution. Otherwise, $t\geq\frac{1}{\e}$, so we have $R^*\leq\frac{\iopt}{t}\leq\e\iopt$ 
and $\lpopt\leq\iopt-\sum_{p=1}^t r_p$. Combining Theorem~\ref{combcost} and
Lemma~\ref{dcost} then yields the theorem.
\end{proofof}

\paragraph{Improved approximation ratio.} 
The improved approximation ratio comes from a better way of combining $F_1$ and $F_2$ in
step B2. The idea is that we can ensure that the dual solutions $\al^1$ and $\al^2$ are
component-wise quite close to each other 
by setting $\dt_z$ in the binary-search procedure to be sufficently small. 
Thus, we may essentially assume that if $T_{1,I}$, $T_{2,I}$ 
denote the tight pairs yielding $F_1$, $F_2$ respectively, then every pair in
$T_{1,I}$ intersects some pair in $T_{2,I}$, because we can augment $T_{2,I}$ to include
non-intersecting pairs of $T_{1,I}$.
This yields dividends when we combine solutions as in step B2, because we can now ensure
that if $\pi(i',r')=(i,r)$, then the pairs of $T_{2,I}$ and $T_{1,I}$ yielding
$(i,r)$ and $(i',r')$ respectively intersect, which yields an improved bound on 
$c_{i,i'}$. 
This yields an improved approximation of $3.83$ for the combination step
(Lemma~\ref{improvcomb}), and hence for the entire algorithm (Theorem~\ref{improvthm});  
we defer the details to Appendix~\ref{append-improved}.  

\begin{theorem}\label{improvthm}
For any $\e > 0$, our algorithm returns a feasible \lbksr-solution of cost at most
$(3.83+O(\e))O^*$ in time $n^{O(1/\e)}$. 
\end{theorem}


\subsection{Approximation algorithm for \lbksro} \label{outl}
We now build upon the ideas in Section~\ref{nonoutl} to devise an $O(1)$-approximation
algorithm for the outlier version \lbksr. 
The high-level approach is similar to the one in Section~\ref{nonoutl}.
We again ``guess'' the $t$ $(i,r)$ pairs $F^O$ corresponding to the facilities with 
largest radii in an optimal solution, 
and consider the modified \kbs-instance $(\D',\cL',k',m)$ (where $\D',\cL',k'$ are defined
as before).
We design a primal-dual algorithm for the Lagrangian relaxation of the
\kbs-problem where we are allowed to pick any number of pairs from $\cL'$ (leaving at
most $m$ uncovered clients) incurring a fixed cost of $\z$ for each pair picked, utilize
this to obtain two solutions $F_1$ and $F_2$, and combine these to extract a low-cost
solution.   
However, the presence of outliers introduces various difficulties both in the primal-dual 
algorithm and in the combination step. 
We consider the following LP-relaxation of the \kbs-problem and its dual (analogous to
\eqref{primal1} and \eqref{dual1}). 

\vspace*{-4ex}

{\centering
\begin{minipage}[t]{0.45\textwidth}
\begin{alignat}{2}
\min & \quad & \sum_{(i,r) \in \cL'} r \cdot y_{i,r} & \tag{P$_2$} \label{primal2} \\
\text{s.t.} && \sum_{(i,r)\in\cL': j\in \cball(i,r)} y_{i,r} & +w_j \geq  1 
\quad \ \ \forall j \in \cD' \notag \\
&& \sum_{(i,r) \in \cL'} y_{i,r} & \le k', \quad \sum_{j\in\D'}w_j\leq m \notag \\[-0.5ex]
&& y,w & \geq 0. \notag
\end{alignat} 
\end{minipage}
\begin{minipage}[t]{0.55\textwidth}
\begin{alignat}{3}
\max & \quad & \sum_{j\in \cD'} \alpha_j - k'\cdot z &-m\cdot\gm & \tag{D$_2$} \label{dual2} \\
\text{s.t.} && \sum_{j\in \cball(i,r)\cap\D'}\alpha_j - z & \leq  r 
\ \ && \forall (i,r)\in\cL' \label{open} \\
&& \al_j & \leq\gm && \forall j\in\cD' \notag \\
&& \al,z,\gm & \geq 0 \notag.
\end{alignat}
\end{minipage}
}

\smallskip
\noindent 
As before, if \eqref{primal2} is infeasible, we reject this guess; so we assume
\eqref{primal2} is feasible in the remainder of this section. 
Let $\lpopt$ denote the optimal value of \eqref{primal2}.
The natural modification of the earlier primal-dual algorithm $\pdalg$ is to now stop the
dual-ascent process when the number of active clients is at most $m$ and set
$\gm=\max_{j\in\D'}\al_j$. 
This introduces the significant complication that 
one may not be able to pay for the $(r+\z)$-cost of non-intersecting tight pairs selected in
the pruning phase by the dual objective value 
$\sum_{j\in\D'}\al_j-m\cdot\gm$, since clients with $\al_j=\gm$ may be needed to pay for
both the $r+\z$-cost of the last tight pair $f=(i_f,r_f)$ but their contribution gets
canceled by the $-m\cdot\gm$ term. 
This issue affects us in various guises. 
First, we no longer obtain an LMP-approximation for the unconstrained problem since
we have to account for the $(r+\z)$-cost of $f$ separately. 
Second, unlike Claim~\ref{fraccost}, given solutions $F_1$ and $F_2$ obtained via binary
search for $\z_1,\z_2\approx\z_1$ respectively with $|F_2|\leq k'<|F_1|$, we now only
obtain a fractional \kbs-solution of cost $O(\lpopt+\z_1)$. 
While one can modify the covering-knapsack-LP based procedure in step B2 of $\kbalg$ to 
combine $F_1$, $F_2$, this only yields a good solution when $\z_1=O(\lpopt)$. The chief
technical difficulty is that $\z_1$ may however be much larger than $\lpopt$. 
Overcoming this obstacle requires various novel ideas and is the key technical
contribution of our algorithm. We design a second combination procedure that is guaranteed
to return a good solution when $\z_1=\Omega(\lpopt)$. This requires establishing certain
structural properties for $F_1$ and $F_2$, using which we argue that one can
find a good solution in the neighborhood of $F_1$ and $F_2$.  

We now detail the changes to the primal-dual algorithm and $\kbalg$ in
Section~\ref{nonoutl} and analyze them to prove Theorem~\ref{kbothm}, which states the 
performance guarantee we obtain for the modified $\kbalg$. As before, for the right guess
of $F^O$, combining this with Lemma~\ref{dcost} immediately yields the following result. 

\begin{theorem} \label{outthm}
There exists a $\bigl(12.365+O(\epsilon)\bigr)$-approximation algorithm for \lbksro
that runs in time $n^{O(1/\epsilon)}$ for any $\e>0$.
\end{theorem}

\paragraph{Modified primal-dual algorithm \boldmath $\pdo(\D',\cL',\z)$.}
This is quite similar to $\pdalg$ (and we again return pairs from $\cL$). 
We stop the dual-ascent process when there are at most $m$ active clients. We set 
$\gamma = \max_{j\in \cD'}\alpha_j$. 
Let $\f=(i_\f,r_\f)$ be the last tight pair added to the tight-pair set $T$, and 
$\cball_\f=\cball(i_\f,r_\f)$. We sometimes abuse notation and use $(i,r)$ to also denote
the singleton set $\{(i,r)\}$. 
For a set $P$ of $(i,r)$ pairs, define $\uncov(P):=\cD'\sm\bigcup_{(i,r)\in P}\cball(i,r)$.
Note that $|\uncov(T\sm\f)|>m\geq|\uncov(T)|$.
Let $\outl$ be a set 
of $m$ clients such that $\uncov(T)\sse\outl\sse\uncov(T\sm\f)$.
Note that $\al_j=\gm$ for all $j\in\outl$.

The pruning phase is similar to before, but we 
only use $\f$ if necessary. Let $T_I$ be a maximal subset of non-intersecting pairs picked
by greedily scanning pairs in $T\sm \f$ in non-increasing order of radius. 
For $i\in\proj(T_I)$, set $\rad(i)$ to be the unique $r$ such that $(i,r)\in T_I$, and 
let $r_i$ be the smallest radius $\rho$ such that $\cball(i,\rho)\supseteq\cball(i',r')$
for every $(i',r')\in T\sm\f$ such that $r'\leq\rad(i)$ and $(i',r')$ intersects
$(i,\rad(i))$. 
Let $F'= \{(i,r_i)\}_{i\in\proj(T_I)}$. 
If $\uncov(F')\leq m$, set $F=F'$.
If $\uncov(F')>m$ and $\exists i\in\proj(F')$ such that $c(i,i_\f)\leq 2R^*$, then
increase $r_i$ so that $\cball(i,r_i)\supseteq\cball_\f$ and let $F$ be 
this updated $F'$.  
Otherwise, set $F=F\cup\f$ and $r_{i_\f}=\rad(i_\f)=r_\f$. 
We return $(F,\f,\outl,\{\rad(i)\}_{i\in\proj(F)},\al,\gm)$.
The proof of Theorem~\ref{thm:outsumr} dovetails the proof of Theorem~\ref{thm:sumr}.

\begin{theorem}\label{thm:outsumr}
Let $(F,\f,\outl,\{\rad(i)\},\al,\gm)=\pdo(\D',\cL',\z)$. 
Then 
\begin{enumerate}[(i), itemsep=0ex, topsep=0.5ex]
\item $\uncov(F)\leq m$, \quad
\item $\cost(F\sm\f)+3|F\sm\f|\z-3R^*\leq 3(\sum_{j\in \cD'}\alpha_j-m\gamma)\leq 3(\lpopt+k'\z)$, 
\item $\bigl\{(i,\rad(i))\bigr\}_{i\in\proj(F)}\sse\cL'$, is a set of non-intersecting
pairs, and $\rad(i)\leq r_i\leq 3R^*\ \forall i\in\proj(F)$, 
\item if $|F\sm\f|\geq k'$ then $\cost(F)\leq 3\cdot\lpopt+4R^*$, and if $|F\sm\f|>k'$
then $\z\leq\lpopt$.
\end{enumerate}
\end{theorem}

\begin{proof}
We first prove parts (i)--(iii).
Let $F'=\{(i,r'_i)\}_{i\in\proj(T_I)}$ be the set of pairs obtained from the set $T_I$ in
the pruning phase. By the same argument as in the proof of Theorem~\ref{thm:sumr}, we have
$r'_i\leq 3\rad(i)\leq 3R^*$ for all $i\in\proj(T_I)$, and $\uncov(F')\sse\uncov(T\sm\f)$. 
If we return $F=F'$, then $|\uncov(F)|\leq m$ by definition.
If $\uncov(F')>m$ and we increase the radius of some $i\in\proj(F')$ with 
$c(i,i_\f)\leq 2R^*$, then we have $r_i\leq\max\{r'_i,3R^*\}\leq 3R^*$ and
$\uncov(F)\sse\uncov(T)$, so $|\uncov(F)|\leq m$. 
If $\f\in F$, then we again have $\uncov(F)\sse\uncov(T)$. This
proves part (i). 
 
The above argument shows that $\cost(F\sm\f)\leq\sum_{i\in\proj(T_I)}3\cdot\rad(i)+3R^*$. 
All pairs in $T_I$ are tight and non-intersecting and $|F\sm\f|=|T_I|$. 
Also, $\outl\sse\uncov(T\sm\f)\sse\uncov(T_I)$. (Recall that $|\outl|=m$ and $\al_j=\gm$
for all $j\in\outl$.) So 
\begin{alignat}{1}
\cost(F\sm\f)+3|F\sm\f|\z-3R^* & \leq\sum_{i\in\proj(T_I)}(3\cdot\rad(i)+3\z)
= \sum_{\substack{i\in\proj(T_I) \\ j\in \cball(i,\rad(i))\cap\D'}}3\alpha_j \notag \\
& \leq 3\Bigl(\sum_{j\in \cD'}\alpha_j-\sum_{j\in\outl}\al_j\Bigr)
= 3\Bigl(\sum_{j\in \cD'}\alpha_j-m\gm\Bigr)
\leq 3(\lpopt+k'\z). \label{eq:outcostS}
\end{alignat}
The last inequality follows since $(\al,\gm,\z)$ is a feasible solution to
\eqref{dual2}. This proves part (ii).

Notice that $\bigl\{(i,\rad(i))\bigr\}_{i\in\proj(F)}$ is $T_I$ if $\f\notin F$, and
$T_I+\f$ otherwise. In the latter case, we know that $c(i,i_\f)>2R^*$ for all
$i\in\proj(T_I)$, so $\f$ does not intersect $(i,\rad(i))$ for any $i\in\proj(T_I)$. 
Thus, all pairs in
$\bigl\{(i,\rad(i))\bigr\}_{i\in\proj(F)}$ are non-intersecting. 
The claim that $\rad(i)\leq r_i$ for all $i\in\proj(F)$ follows from 
exactly the same argument as that in the proof of Theorem~\ref{thm:sumr}. 

Part (iv) follows from part (ii) and \eqref{eq:outcostS}. The bound on $\cost(F)$ follows 
from part (ii) since that $\cost(F)\leq\cost(F\sm\f)+R^*$. Inequality \eqref{eq:outcostS}
implies that $|F\sm\f|\z\leq\sum_{i\in\proj(T_I)}(\rad(i)+\z)\leq\lpopt+k'\z$, and so
$\z\leq\lpopt$ if $|F\sm\f|>k'$.
\end{proof}

\paragraph{Modified algorithm \boldmath $\kbo(\D',\cL',k',\e)$.} 
As before, we use binary search to find solutions $F_1$, $F_2$ and extract a low-cost
solution from these. 
The only changes to step B1 are as follows.
We start with $\z_1=0$ and $\z_2=2nk'c_{\max}$; 
for this $\z_2$, we argue below that $\pdo$ returns at most $k'$ pairs. We stop when 
$\z_2-\z_1\leq\delta_\z:=\frac{\e\lpopt}{3n2^n}$.  
We {\em do not stop} even if $\pdo$ returns a solution $(F,\ldots)$ 
with $|F|=k'$ for some $\z=\frac{\z_1+\z_2}{2}$, since 
{\em Theorem~\ref{thm:outsumr} is not strong enough to bound $\cost(F)$ even when this
happens!}. 
If $|F|>k'$, we update $\z_1\assign\z$ and the $F_1$-solution; otherwise, we update
$\z_2\assign\z$ and the $F_2$-solution. Thus, we maintain 
that $k_1=|F_1|>k'$, and $k_2=|F_2|\leq k'$.

\begin{claim} \label{z2pairs}
When $\z=\z_2=2nk'c_{\max}$, $\pdo$ returns at most $k'$ pairs. 
\end{claim}

\begin{proof}
Let $(F,f,\out,\{\rad(i)\}_{i\in\mu(F)},\al,\gm)$ be the output of $\pdo$ for
this $\z$. Let $T$ be the sight of tight pairs after the dual-ascent process. 
Observe that $\gm\geq 2k'c_{\max}$, since for any tight pair $(i,r)\in T$, we have that   
$n\gm\geq\sum_{j\in\cball(i,r)\cap\D'}\al_j\geq\z$.
We have $\sum_{j\in\D'}\al_j-m\gm\leq\lpopt+k'\z\leq k'c_{\max}+k'\z$.
On the other hand, since $\uncov(T\sm f)\sm\out\neq\es$ and $\al_j=\gm$ for all
$j\in\uncov(T\sm f)$, we also have the lower bound
$$
\sum_{j\in\D'}\al_j-m\gm\geq
\sum_{\substack{i\in\mu(F\setminus f) \\ j\in\cball(i,\rad(i))\cap\D'}}\al_j+\gm
\geq|F\sm f|\z+\gm.
$$
So if $|F|>k'$, we arrive at the contradiction that $\gm\leq k'c_{\max}$. 
\end{proof}

The main change is in the way solutions $F_1$, $F_2$ are combined. We adapt step B2 to
handle outliers (procedure $\A$ in Section~\ref{aroutine}), but the key extra ingredient
is that we devise an alternate combination procedure $\B$ (Section~\ref{broutine}) that 
returns a low-cost solution when $z_1=\Om(\lpopt)$. We return the better of the
solutions output by the two procedures. We summarize these changes at the end in Algorithm 
$\kbo(\D',\cL',k',\e)$ and state the approximation bound for $\kbo$ (Theorem~\ref{kbothm}). 
Combining this with Lemma~\ref{dcost} (for the right selection of $t$ $(i,r$) pairs)
immediately yields Theorem~\ref{outthm}.  

We require the following {\em continuity lemma}, which is essentially Lemma 6.6
in~\cite{charikar2001clustering}; 
we include a proof in Appendix~\ref{append-charikar} for completeness. 

\begin{lemma}\label{lem:charikaralpha}
Let $(F_p,\ldots,\al^p,\gm^p)=\pdo(\D',\cL',\z_p)$ for $p=1,2$, where 
$0\leq \z_2-\z_1\leq\dt_z$.
Then, 
$\|\al_j^1-\al_j^2\|_\infty\leq 2^n\delta_\z$ and $|\gm^1-\gm^2|\leq 2^n\dt_z$. 
Thus, if \eqref{open} 
is tight for some $(i,r)\in\cL'$ in one execution, then 
$\sum_{j\in\cball(i,r)\cap\D'} \alpha^p_j \geq r+\z_1-2^n\delta_{\z}$ for $p=1,2$.
\end{lemma}

\subsubsection{Combination subroutine $\A\bigl((F_1,\rad_1),(F_2,\rad_2)\bigr)$} 
\label{aroutine} 
As in step B2, we cluster the $F_1$-pairs around $F_2$-pairs in stars. However,
unlike before, some $(i',r')\in F_1$ may remain {\em unclustered} and
and we may not pick $(i',r')$ or some pair close to it. 
Since we do not cover all clients covered by $F_1$, we need to cover a suitable 
number of clients from $\uncov(F_1)$. We again setup an LP to obtain a suitable collection 
of pairs. Let $\uc_p$ denote $\uncov(F_p)$ and $\D_p:=\D'\sm\uc_p$ for $p=1,2$. 
Let $\pi : F_1\rightarrow F_2 \cup \{\emptyset\}$ be defined as follows:
for each $(i',r')\in F_1$, 
if $(i',r')\in F_1$ intersects some $F_2$-pair, pick such an intersecting $(i,r)\in F_2$
and set $\pi(i',r')=(i,r)$; otherwise, set $\pi(i',r')=\es$. In the latter case,
$(i',r')$ is unclustered, and $\cball(i',r')\sse\uc_2$.
Define $\nstar_{i,r}=\pi^{-1}(i,r)$ for all $(i,r)\in F_2$.
Let $\Qc=\pi^{-1}(\es)$. 
Let $\{\uc_1(i,r)\}_{(i,r)\in F_2}$ be a partition of $\uc_1\cap\D_2$ such that
$\uc_1(i,r)\sse\uc_1\cap\cball(i,r)$ for all $(i,r)\in F_2$. 
Similarly, let $\{\uc_2(i',r')\}_{(i',r')\in F_1}$ be a partition of $\uc_2\cap\D_1$ such that
$\uc_2(i',r')\sse\uc_2\cap\cball(i',r')$ for all $(i',r')\in F_1$. 
We consider the following 2-dimensional covering knapsack LP. 
\begin{alignat}{2}
\negthickspace \min & \quad & \sum_{(i,r)\in F_2}\Bigl(x_{i,r}(2r+
{\textstyle \sum_{(i',r')\in\nstar_{i,r}}}2r') +
(1-x_{i,r})&{\textstyle \sum_{(i',r')\in\nstar_{i,r}}}r'\Bigr) 
+\sum_{(i',r')\in\Qc}\negthickspace q_{i',r'}\cdot r' \tag{2C-P} \label{newcklp} \\
\text{s.t.} & \quad & \sum_{(i,r)\in F_2}\bigl(x_{i,r}+|\nstar_{i,r}|(1-x_{i,r})\bigr) &
+\sum_{(i',r')\in\Qc}q_{i',r'} \leq k \label{eq:sizebound2} \\
&& \sum_{(i,r)\in F_2}(1-x_{i,r})|\uc_1(i,r)|
+\sum_{(i',r')\in\Qc}(1-q_{i',r'}) & |\uc_2(i',r')| \leq m-|\uc_1\cap\uc_2| 
\label{eq:outbound} \\
&& 0 \leq x_{i,r} \leq 1 \quad \forall (i,r)\in F_2, & \qquad 
0\leq q_{i',r'}\leq 1 \quad \forall (i',r')\in\Qc. \notag
\end{alignat}
The interpretation of the variable $x_{i,r}$ is similar to before.
If $x_{i,r}=0$, or $x_{i,r}=1,\ \nstar_{i,r}\neq\es$, we proceed as in step B2 (i.e.,
select all pairs in $\nstar_{i,r}$, or pick some $(i',r')\in\nstar_{i,r}$ and expand its
radius suitably). 
But if $x_{i,r}=1,\ \nstar_{i,r}=\es$, then we may also pick $(i,r)$ (see
Theorem~\ref{combA}).  
Variable $q_{i',r'}$ indicates if we pick $(i',r')\in F_1$. 
The number of uncovered clients in such a solution is at most 
$|\uc_1\cap\uc_2|$ + (LHS of \eqref{eq:outbound}), and \eqref{eq:outbound} enforces
that this is at most $m$. 

Let $(x^*,q^*)$ be an extreme-point optimal solution to \eqref{newcklp}. The number of
fractional components in $(x^*,q^*)$ is at most the number of tight constraints from
\eqref{eq:sizebound2}, \eqref{eq:outbound}. 
We exploit this to round $(x^*,q^*)$ to an integer solution $(\tx,\tq)$ of good objective
value (Lemma~\ref{combAlem2}), and then use $(\tx,\tq)$ to extract a good set of
pairs as sketched above (Theorem~\ref{combA}). 
Recall that $k_1=|F_1|$, $k_2=|F_2|$.
Let $a,b\geq 0$ be such that $ak_1+bk_2=k'$, $a+b=1$. 
{Let $C_1=\cost(F_1)$ and $C_2=\cost(F_2)$.} 

\begin{lemma} \label{combAlem1}
The following hold.
\begin{enumerate}[(i), itemsep=0ex, topsep=0.5ex]
\item $aC_1+bC_2\leq(3+\e)\lpopt+4R^*+3\z_1$, 
\item $\OPT_{\text{\ref{newcklp}}}\leq 2bC_2+(1+b)C_1$.
\end{enumerate}
\end{lemma}

\begin{proof}
Part (i) follows easily from part (ii) of Theorem~\ref{thm:outsumr} and since
$\cost(F_p)\leq\cost(F_p\sm\f_p)+R^*$ for $p=1,2$. So we have 
$C_1+3(k_1-1)\z_1\leq 3(\lpopt+k'\z_1)+4R^*$ and 
$C_2+3(k_2-1)\z_2\leq 3(\lpopt+k'\z_2)+4R^*$. Combining these, we obtain
\begin{equation*}
\begin{split}
aC_1+bC_2 & \leq 3\lpopt+3k'(a\z_1+b\z_2)-3(ak_1\z_1+bk_2\z_2)+3(a\z_1+b\z_2)+4R^* \\
& \leq 3(\lpopt+k'\z_2)-3k'\z_2+3ak_1\dt_z+3\z_1+3b\dt_z+4R^* \\ 
& \leq (3+\e)\lpopt+4R^*+3\z_1.
\end{split}
\end{equation*}
The second inequality follows since $0\leq\z_2-z_1\leq\dt_z$.

For part (ii), we claim that setting $x_{i,r}=b$ for all $(i,r)\in F_2$, and $q_{i',r'}=a$
for all $(i',r')\in\Qc$ yields a feasible solution to \eqref{newcklp}.
The LHS of \eqref{eq:sizebound2} evaluates to $ak_1+bk_2$, which is exactly $k'$.
The first term on the LHS of \eqref{eq:outbound} evaluates to 
$a\sum_{(i,r)\in F_2}|\uc_1(i,r)|=a|\uc_1\cap\D_2|=a|\uc_1\sm\uc_2|$ since
$\{\uc_1(i,r)\}_{(i,r)\in F_2}$ is a partition of $\uc_1\cap\D_2$.  
Similarly, the second term on the LHS of \eqref{eq:outbound} evaluates to at most
$b|\uc_2\cap\D_1|=b|\uc_2\sm\uc_1|$. 
So we have 
$$
\text{(LHS of \eqref{eq:outbound})}+|\uc_1\cap\uc_2|=a|\uc_1|+b|\uc_2|\leq m
$$
since $|\uc_1|,|\uc_2|\leq m$.
The objective value of this solution is $2bC_2+2bC_1+(1-b)C_1=2bC_2+(1+b)C_1$.   
\end{proof}

Let $\Pc=\{(i,r)\in F_2:\nstar_{i,r}=\es\}$.

\begin{lemma} \label{combAlem2}
$(x^*,q^*)$ can be rounded to a feasible integer solution $(\tx,\tq)$ to
\eqref{newcklp} of objective value at most $\OPT_{\text{\ref{newcklp}}}+O(R^*)$.
\end{lemma}

\begin{proof}
Let $S$ be the set of fractional components of $(x^*,q^*)$.
As noted earlier, 
$|S|$ is at most the number of tight constraints from \eqref{eq:sizebound2},
\eqref{eq:outbound}. Let 
$$
l^*:=\sum_{(i,r)\in S\cap F_2}\bigl(x^*_{i,r}+|\nstar_{i,r}|(1-x^*_{i,r})\bigr)
+\sum_{(i',r')\in S\cap\Qc}q^*_{i',r'} 
$$
denote the contribution of the fractional components of $(x^*,q^*)$ to the LHS of
\eqref{eq:sizebound2}. Note that if \eqref{eq:sizebound2} is tight, then $l^*$ must be an
integer. For a vector $v=(v_j)_{j\in I}$ where $I$ is some index-set, let $\ceil{v}$
denote $\bigl(\ceil{v_j}\bigr)_{j\in I}$. 
We round $(x^*,q^*)$ as follows.
\begin{itemize}[nosep]
\item If $l^*\geq 2$ or $|S|\leq 1$ or $|S\cap(F_2\sm\Pc)|\geq 1$, set
$(\tx,\tq)=\ceil{(x^*,q^*)}$.  

\item Otherwise, we set $\tx_{i,r}=x^*_{i,r}$, $\tq_{i',r'}=q^*_{i',r'}$ for all the
integer-valued coordinates. We set the fractional component with larger absolute
coefficient value on the LHS of \eqref{eq:outbound} equal to 1 
and the other fractional component to 0. 
\end{itemize}
We prove that $(\tx,\tq)$ is a feasible solution to \eqref{newcklp}.
Note that \eqref{eq:outbound} holds for $(\tx,\tq)$ since 
we always have
\begin{equation*}
\begin{split}
\sum_{(i,r)\in F_2}(1-\tx_{i,r})|\uc_1(i,r)|&+\sum_{(i',r')\in\Qc}(1-\tq_{i',r'})|\uc_2(i',r')|
\\
& \leq \sum_{(i,r)\in F_2}(1-x^*_{i,r})|\uc_1(i,r)|+\sum_{(i',r')\in\Qc}(1-q^*_{i',r'})|\uc_2(i',r')|.
\end{split}
\end{equation*}
Clearly, the contribution to the LHS of \eqref{eq:sizebound2} from the components not in
$S$ is the same in both $(\tx,\tq)$ and $(x^*,q^*)$. 
Let $l$ denote the contribution from $(\tx,\tq)$ to the LHS of \eqref{eq:sizebound2} from
the components in $S$. Clearly, $l$ is an integer.

If $l^*\geq 2$, then $l=2$. If $|S|\leq 1$,
then $l=1$. If $l^*\geq 1$, then in these cases the LHS of \eqref{eq:sizebound2} evaluated
at $(\tx,\tq)$ is at most the LHS of \eqref{eq:sizebound2} evaluated at $(x^*,q^*)$.
If $l^*<1$ and $|S|\leq 1$ (so $l=1$), then since $l^*$ is fractional, we know that
\eqref{eq:sizebound2} is not tight for $(x^*,q^*)$. So despite the increase in LHS of
\eqref{eq:sizebound2}, we have that \eqref{eq:sizebound2} holds for $(\tx,\tq)$. 
If $|S|=2$ and $|S\cap(F_2\sm\Pc)|\geq 1$, then we actually have $l^*>1$
and $l=2$. Again, since $l^*$ is fractional, we can conclude that $(\tx,\tq)$ satisfies
\eqref{eq:sizebound2} despite the increase in LHS of \eqref{eq:sizebound2}.
Finally, suppose $l^*<2$, $|S|=2$, and $S\cap (F_2\sm\Pc)=\es$. 
Then the contribution from $S$ to the LHS of \eqref{eq:sizebound2} is 
$\sum_{(i,r)\in S\cap F_2}x_{i,r}+\sum_{(i',r')\in S\cap\Qc}q_{i',r'}$, and at most one of
the components in $S$ is set to 1 in $(\tx,\tq)$. So $l=1$, and either $l\leq l^*$ or
$l^*<1$, and in both cases \eqref{eq:sizebound2} holds for $(\tx,\tq)$.

To bound the objective value of $(\tx,\tq)$, notice that
compared to $(x^*,q^*)$, the solution $(\tx,\tq)$ pays extra only for the components that
are rounded up. There are at most two such components, and their objective-function
coefficients are bounded by $15R^*$, so the objective value of $(\tx,\tq)$ is at most
$\OPT_{\text{\ref{newcklp}}}+30R^*$. 
\end{proof}

\begin{theorem} \label{combA}
The integer solution $(\tx,\tq)$ returned by Lemma~\ref{combAlem2} yields a solution 
$\bigl(F,\{\rad(i)\}_{i\in\proj(F)}\bigr)$ to the \kbs-problem 
with $\cost(F)\leq\bigl(6.1821+O(\e)\bigr)(\lpopt+\z_1)+O(R^*)$ where 
$\bigl\{(i,\rad(i))\bigr\}_{i\in\proj(F)}\sse\cL'$ is a set of non-intersecting pairs. 
\end{theorem}

\begin{proof}
Unlike in step B2 of \kbalg, we will not simply pick a subset of pairs of $F_1$ and expand
their radii. We will sometimes need to pick pairs from $F_2$ in order to ensure that we
have at most $m$ outliers, but we need to be careful in doing so because we also need to
find suitable radii for the facilities we pick so that we obtain non-intersecting pairs.

We first construct $F''$ as follows.
If $\tq_{i',r'}=1$, we include $(i',r')\in F''$ and set $\rad(i')=\rad_1(i')$.
If $\tx_{i,r}=0$, we include all pairs in $\nstar_{i,r}$ in $F''$ and set
$\rad(i')=\rad_1(i')$ for all $(i',r')\in\nstar_{i,r}$.
If $\tx_{i,r}=1$ and $\nstar_{i,r}\neq\es$, we pick a pair in $(i',r')\in\nstar_{i,r}$, and
include $(i',2r+r'+\max_{(i'',r'')\in\nstar_{i,r}\sm\{(i',r')\}}2r'')$ in $F''$. We set
$\rad(i')=\rad_1(i')$.  
Now we initialize $F'=F''$ and consider all $(i,r)\in\Pc$ with $\tx_{i,r}=1$.
If $(i,r)$ does not intersect any $(i',r')\in F''$ 
then we add $(i,r)$ to $F'$, and set $\rad(i)=\rad_2(i)$. 
Otherwise, if $(i,r)$ intersects some $(i',r')\in F''$, then
we replace $(i',r')\in F'$ with $(i',r'+2r)$.
We have thus ensured that $\bigl\{(i,\rad(i))\bigr\}_{i\in\proj(F')}\sse\cL'$ and 
consists of non-intersecting pairs.
Note that in all the cases above, the total cost of the pairs we include when we process
some $\tq_{i',r'}$ or $\tx_{i,r}$ term is at most the total contribution to the objective
function from the $\tq_{i',r'}$ term, or the $\tx_{i,r}$ and $1-\tx_{i,r}$ terms.
Therefore, $\cost(F')$ is at most the objective value of $(\tx,\tq)$. Finally, we argue
that $|\uncov(F')|\leq m$. 
We have 
$|\uncov(F')|\leq|\uc_1\cap\uc_2|+|\uncov(F')\cap\D_1|+|\uncov(F')\cap\D_2\cap\uc_1|$.
Observe that for every client $j\in\uncov(F')\cap\D_1$ and every $(i',r')\in F_1$ such
that $j\in\cball(i',r')$, it must be that $(i',r')\in\Qc$ and $\tq_{i',r'}=0$. It follows
that $j\in\uc_2(i',r')$ for some $(i',r')\in\Qc$ with $\tq_{i',r'}=0$. Therefore,  
$|\uncov(F')\cap\D_1|\leq\sum_{(i',r')\in\Qc}(1-\tq_{i',r'})|\uc_2(i',r')|$. 
Similarly, for every $j\in\uncov(F')\cap\D_2\cap\uc_1$ and every $(i,r)\in F_2$
such that $j\in\cball(i,r)$, we must have $(i,r)\in\Pc$ and $\tx_{i,r}=0$; hence,
$j\in \uc_1(i,r)$ for some $(i,r)\in\Pc$ with $\tx_{i,r}=0$.
Therefore,
$|\uncov(F')\cap\D_2\cap\uc_1|\leq\sum_{(i,r)\in\Pc}(1-\tx_{i,r})|\uc_1(i,r)|$. Thus, since
$(\tx,\tq)$ is feasible, constraint \eqref{eq:outbound} implies that 
$|\uncov(F')|\leq m$. 

We return $(F_2,\rad_2)$ if $\cost(F_2)\leq\cost(F')$, and
$\bigl(F',\{\rad(i)\}_{i\in\proj(F')}\bigr)$ otherwise. Combining the above bound on
$\cost(F')$ with part (ii) of Lemma~\ref{combAlem1} and Lemma~\ref{combAlem2}, we obtain 
that the cost of the solution returned is at most 
\begin{equation*}
\begin{split}
\min\bigl\{C_2,2bC_2&+(1+b)C_1\bigr\}+30R^*\leq 2.0607\bigl(aC_1+bC_2\bigr)+30R^* \\
& \leq 2.0607\Bigl((3+\e)\lpopt+4R^*+3\z_1\Bigr)+30R^*
\leq (6.1821+3\e)(\lpopt+\z_1)+39R^*.
\end{split}
\end{equation*}
The first inequality follows from Claim~\ref{clm:costab}, and the second
follows from part (i) of Lemma~\ref{combAlem1}.
\end{proof}

\subsubsection{Subroutine $\B\bigl((F_1,\f_1,\outl_1,\rad_1,\al^1,\gm^1),(F_2,\f_2,\outl_2,\rad_2,\al^2,\gm^2)\bigr)$} 
\label{broutine}
Subroutine $\A$ in the previous section yields
a low-cost solution only if $\z_1=O(\lpopt)$. We complement subroutine $\A$ by now
describing a procedure that returns a good solution  
when $\z_1$ is large. We assume in this section that $\z_1>(1+\e)\lpopt$.
Then 
$|F_1\sm\f_1|\leq k'$ (otherwise $\z_1\leq\lpopt$ by part (iv) of
Theorem~\ref{thm:outsumr}), 
so $|F_1\sm\f_1|\leq k'<|F_1|$, which means that $k_1=k'+1$ and $\f_1\in F_1$. Hence,
$\al^1_j=\gm^1$ for all $j\in\cball_{\f_1}\cap\D'$. 

First, we take care of some simple cases. 
If there exists $(i,r)\in F_1\sm\f_1$ such that
$|\uncov\bigl(F_1\sm\{\f_1,(i,r)\}\cup (i,r+12R^*)\bigr)|\leq m$, then 
set $F=F_1\sm\{\f_1,(i,r)\}\cup (i,r+12R^*)$. 
We have $\cost(F)=\cost(F_1\sm\f_1)+12R^*\leq 3\cdot\lpopt+15R^*$ 
(by part (ii) of Theorem~\ref{thm:outsumr}). 
If there exist pairs $(i,r), (i',r')\in F_1$ such that $c(i,i')\leq 12 R^*$, take 
$r''$ to be the minimum $\rho\geq r$ such that $\cball(i',r')\sse\cball(i,\rho)$ and
set $F=F_1\sm\{(i,r),(i',r')\}\cup (i,r'')$. We have $\cost(F)\leq\cost(F_1\sm\f_1)+13R^*\leq 3\cdot\lpopt+16R^*$.
In both cases, we return $\bigl(F,\{\rad_1(i)\}_{i\in\proj(F)}\bigr)$.

So we assume in the sequel that neither of the above apply. 
In particular, all pairs in $F_1$ are well-separated.
Let $\AT=\{(i,r)\in\cL': \sum_{j\in\cball(i,r)\cap\D'} \alpha^1_j \geq r+\z_1-2^n\delta_z\}$ 
and  
$\AD = \{j\in\D': \alpha^1_j \geq \gamma^1 - 2^n\delta_z\}$. 
By Lemma~\ref{lem:charikaralpha}, $\AT$ includes the tight pairs of
$\pdo(\D',\cL',\z_p)$ for both $p=1,2$, and $\outl_1\cup\outl_2 \sse\AD$. 
Since the tight pairs $T_2$ used for building solution $F_2$ are almost tight in
$(\alpha^1, \gamma^1, \z_1)$, we swap them in and swap out pairs from $F_1$ one by
one while maintaining a feasible solution. 
Either at some point, we will be able to remove $\f$, which will give us a solution of
size $k'$, or we will obtain a bound on $\cost(F_2)$. The following lemma is our main tool
for bounding the cost of the solution returned.

\begin{lemma}\label{lem:size>kAD}
Let $F\sse\cL'$, and let $T_F=\bigl\{(i,r'_i)\bigr\}_{i\in\proj(F)}$ where 
$r'_i\leq r$ for each $(i,r)\in F$. Suppose $T_F\sse\AT$ and pairs in $T_F$ are
non-intersecting. If $|F|\geq k'$ and $|\AD\sm\bigcup_{(i,r)\in F}\cball(i,r))| \geq m$ 
then $\cost(T_F)\leq(1+\e)\lpopt$.
Moreover, if $|F|>k'$ then $\z_1\leq(1+\e)\lpopt$.
\end{lemma}

\begin{proof}
Let $\outl_F$ be a subset of exactly $m$ of clients from 
$\AD\sm\bigcup_{(i,r)\in F}\cball(i,r)$. Since the pairs in $T_F$ are non-intersecting and
almost tight, 
$\sum_{i\in\proj(F)}(r'_i+\z_1) \leq \sum_{j\in \cD'\sm\outl_F}(\alpha^1_j+2^n\delta_z)$, so   
\begin{equation*}
\sum_{i\in\proj(F)} (r'_i+\z_1) 
\leq \sum_{j\in \cD'}(\alpha^1_j + 2^n\delta_z)-m(\gamma^1-2^n\delta_z) 
\leq \sum_{j\in \cD'} \alpha^1_j -m\gamma^1 + (m+|\cD'|)2^n \delta_z 
\leq (1+\e)\lpopt+k'\z_1
\end{equation*}
where the last inequality follows since $(\al^1,\gm^1,\z_1)$ is a feasible solution to
\eqref{dual2}. 
So $\cost(T_F)\leq (1+\epsilon)\lpopt$ if $|T_F|=|F|\geq k'$, and
$\z_1\leq(1+\epsilon)\lpopt$ if $|F|>k'$.
\end{proof}

Define a mapping $\psi:F_2\rightarrow F_1\sm \f_1$  as follows. 
Note that any $(i,r)\in F_2$ may intersect with at most one $F_1$-pair: 
if it intersects $(i',r'),(i'',r'')\in F_1$, then we have $c(i',i'')\leq 12R^*$. 
First, for each $(i,r)\in F_2$ that intersects with some $(i',r')\in F_1$, we set
$\psi(i,r)=(i',r')$. Let $M\sse F_2$ be the $F_2$-pairs mapped by $\psi$ this way. 
For every $(i,r)\in F_2\sm M$, 
we arbitrarily match $(i,r)$ with a {\em distinct} $(i',r')\in F_1\sm\psi(M)$.
We claim that $\psi$ is in fact a one-one function.

\begin{lemma} \label{onetoone}
Every $(i,r)\in F_1\sm \f_1$ intersects with at most one $F_2$-pair.
\end{lemma}

\begin{proof}
Suppose two pairs $(i_1,r_1), (i_2,r_2) \in F_2$ intersect with a common pair 
$(i,r)\in F_1\sm\f_1$. Let $T_{1,I}$ be the tight pairs corresponding to $F_1\sm\f_1$
obtained from (the pruning phase of) $\pdo(\D',\cL',\z_1)$. 
Let $(i,\rad_1(i))\in T_{1,I}$ be the tight pair corresponding to $(i,r)$. 
Let $(i_1,\rad_2(i_1)), (i_2,\rad_2(i_2))$ be the tight pairs corresponding to
$(i_1,r_1), (i_2,r_2)$ obtained from $\pdo(\D',\cL',\z_2)$.
Let $F''=F_1\sm\{\f_1,(i,r)\}\cup(i,r+12R^*)$.
We show that either $\z_1\leq\lpopt$ or $|\uncov(F'')|\leq m$, both of which
lead to a contradiction.

Define $F'=F_1\sm\{\f_1,(i,r)\}\cup\{(i_1,r_1),(i_2,r_2)\}$, so $|F'|=k+1$. Consider the
set $T_{F'}= T_{1,I}\sm\{(i,\rad_1(i))\}\cup\{(i_1,\rad_2(i_1)),(i_2,\rad_2(i_2))\}$. 
Since $(i_1,\rad_2(i_1)$ and $(i_2,\rad_2(i_2))$ are non-intersecting and they do not
intersect with any pair in $T_{1,I}\sm(i,\rad_1(i))$, the pairs in $T_{F'}$ are
non-intersecting. Also, $T_{F'}\sse\AT$. 
If $|\AD\cap\uncov(F')|=|\AD\sm\bigcup_{(i',r')\in F'}\cball(i',r')|\geq m$, then
$\z_1\leq\lpopt$ by Lemma~\ref{lem:size>kAD}. 

Otherwise, note that every client in $\cball(i_1,r_1)\cup\cball(i_2,r_2)$ is at distance
at most $r+2\max\{r_1,r_2\}\leq r+6R^*$ from $i$. So we have
$\uncov(F'')\sse\uncov(F)\cup\cball_{\f_1}\sse\AD$ and $\uncov(F'')\sse\uncov(F')$. So
$|\uncov(F'')|\leq|\AD\cap\uncov(F')|\leq m$.
\end{proof}

Let $F'_2$ be the pairs $(i,r)\in F_2$ such that if $(i',r')= \psi(i,r)$, then $r'<r$.
Let $P=F'_2\cap M$ and $Q=F'_2\sm M$. For every $(i',r')\in\psi(Q)$ and
$j\in\cball(i',r')$, we have $j\in\uncov(F_2)\sse\AD$ (else $(i',r')$ would lie in
$\psi(M)$). 
Starting with $F=F_1\sm\f_1$, we iterate over $(i,r)\in F'_2$ and do the following. Let
$(i',r')=\psi(i,r)$. If $(i,r)\in P$, we update $F\assign F\sm(i',r')\cup(i,r+2r')$ (so
$\cball(i,r+2r')\supseteq\cball(i',r')$),
else we update $F\assign F\sm(i',r')\cup(i,r)$. 
Let $T_F=\{(i,\rad_1(i))\}_{(i,r)\in F\cap F_1}\cup\{(i,\rad_2(i))\}_{(i,r)\in F\sm F_1}$.
Note that $|F|=k'$ and $\uncov(F)\sse\AD$ at all times.
Also, since $(i,r)$ intersects only $(i',r')$, which we remove when $(i,r)$ is added, 
we maintain that $T_F$ is a collection of non-intersecting pairs and a subset of
$\AT\sse\cL'$. This process continues until $|\uncov(F)|\leq m$, 
or when all pairs of $F'_2$ are swapped in. 
In the former case, we argue that $\cost(F)$ is small and return 
$\bigl(F,\{\rad_1(i)\}_{(i,r)\in F\cap F_1}\cup\{\rad_2(i)\}_{(i,r)\in F\sm F_1}\bigr)$.  
In the latter case, we show that $\cost(F'_2)$, and hence $\cost(F_2)$ is
small, and return $(F_2,\rad_2)$. 

\begin{lemma} \label{lem:uncov} \label{lem:forterm} \label{bapprox}
(i) If the algorithm stops with $|\uncov(F)|\leq m$, then 
$\cost(F)\leq (9+3\epsilon)\lpopt + 18 R^*$. 

\noindent 
(ii) If case (i) does not apply, then $\cost(F_2) \leq (3+3\epsilon)\lpopt+9R^*$.

\noindent
(iii) The pairs corresponding to the radii returned are non-intersecting
and form a subset of $\cL'$.
\end{lemma}

\begin{proof}
Part (iii) follows readily from the algorithm description and the discussion above.
Consider part (i).
Let $(i,r)\in F'_2$ be the last pair scanned by the algorithm before it terminates, and
$(i',r')=\psi(i,r)$. 
Let $F'$ be the set $F$ before the last iteration. So $F'=F\sm (i,r+2r')\cup(i',r')$ if
$(i,r)\in P$, and $F'=F\sm(i,r)\cup(i',r')$ if $(i,r)\in Q$.
Note that $r+2r'\leq 9R^*$.
Since $\uncov(F')\sse\AD$ and $|\uncov(F')|>m$, by Lemma~\ref{lem:size>kAD}, 
we have $\cost(T_{F'})\leq(1+\e)\lpopt$. 
For all $(i,r)\in F_1$, we have $r\leq 3\rad_1(i)$ (since $\f_1\in F_1$).
For all but at most one $(i,r)\in F_2$, we have $r\leq 3\rad_2(i)$ and for the one
possible exception, we have $r\leq 3R^*$.
Therefore, 
\begin{equation*}
\begin{split}
\cost(F) & \leq\cost(F'\cap F_1)+\cost(F'\sm F_1)+9R^*
\leq 3\cdot\cost(T_{F'})+3R^*+2\cdot\cost(F_1\sm F')+9R^* \\
& \leq 3(1+\e)\lpopt+3R^*+2(3\cdot\lpopt+3R^*)+9R^*=\bigl(9+3\e)\lpopt+18R^*.
\end{split}
\end{equation*}
The second inequality above follows since 
$\cost(F'\cap F_1)\leq\sum_{(i,r)\in F'\cap F_1}3\rad_1(i)$ and 
$\cost(F'\sm F_1)\leq\sum_{(i,r)\in F'\sm F_1}3\rad_2(i)+3R^*+2\cost(F_1\sm F')$.

For part (ii), Lemma~\ref{lem:size>kAD} shows that $\cost(T_F)\leq(1+\e)\lpopt$, and so 
$\cost(F'_2)+\cost\bigl(F_1\sm(\f_1\cup\psi(F'_2))\bigr)\leq 3\cdot\cost(T_F)+3R^*$. 
Now 
\begin{equation*}
\begin{split}
\cost(F_2)=\cost(F'_2)+\cost(F_2\sm F'_2)
& \leq \cost(F'_2)+\cost\bigl(\psi(F_2\sm F'_2)\bigr) \\
& = \cost(F'_2)+\cost\bigl(F_1\sm(\f_1\cup\psi(F'_2)\bigr)
\leq 3(1+\e)\cdot\lpopt+3R^*
\end{split}
\end{equation*}
where the first inequality follows by the definition of $F'_2$.
\end{proof}

{\small 
\newpage \hrule
\begin{kballoalg}
Output: $F\sse\cL$ with $|F|\leq k'$, a radius $\rad(i)$ for all $i\in\proj(F)$.
\begin{enumerate}[label=C\arabic*., topsep=0.5ex, itemsep=0ex, labelwidth=\widthof{C3.},
    leftmargin=!] 
\item {\bf Binary search.\ }
Let $(F_1,\rad_1,\ldots)=\pdo(\D',\cL',0)$. If $|F_1|\leq k'$ pairs, return
$(F_1,\rad_1)$. 
Else perform binary-search in the range $[0,nc_{\max}]$ to find $\z_1,\z_2$ with
$0\leq\z_2-\z_1\leq\dt_z=\frac{\e\lpopt}{3n2^n}$ such that 
letting $(F_p,\f_p,\outl_p,\rad_p,\al^p,\gm^p)=\pdalg(\D',\cL',\z_p)$ for $p=1,2$, we have
$|F_2|\leq k'<|F_1|$.

\item 
Let $\bigl(F_\A,\{\rad_\A(i)\}_{i\in\proj(F_\A)}\bigr)=\A\bigl((F_1,\rad_1),(F_2,\rad_2)\bigr)$  
(Section~\ref{aroutine}). 
If $|F_1\sm\f_1|>k'$, return $(F_\A,\rad_\A)$.

\item If $\exists (i,r)\in F_1\sm\f_1$ such that
$|\uncov\bigl(F_1\sm\{\f_1,(i,r)\}\cup (i,r+12R^*)\bigr)|\leq m$, then 
set $F=F_1\sm\{\f_1,(i,r)\}\cup (i,r+12R^*)$. 
If $\exists (i,r), (i',r')\in F_1$ such that $c(i,i')\leq 12 R^*$, 
let $r''$ be the minimum $\rho\geq r$ such that $\cball(i',r')\sse\cball(i,\rho)$;
set $F=F_1\sm\{(i,r),(i',r')\}\cup (i,r'')$. 
If either of the above apply, return $\bigl(F,\{\rad_1(i)\}_{i\in\proj(F)}\bigr)$.

\item Let $\bigl(F_\B,\{\rad_\B(i)\}_{i\in\proj(F_\B)}\bigr)$ be the output of subroutine
$\B$ (Section~\ref{broutine}). 

\item If $\cost(F_\A)\leq\cost(F_\B)$, return $(F_\A,\rad_\A)$, else return
$(F_\B,\rad_\B)$.  
\end{enumerate}
\end{kballoalg}
\hrule
}

\begin{theorem} \label{kbothm}
$\kbo(\D',\cL',k')$ returns a solution $(F,\rad)$ with
$\cost(F)\leq\bigl(12.365+O(\e)\bigr)\cdot\lpopt+O(R^*)$ where
$\bigl\{(i,\rad(i))\bigr\}_{i\in\proj(F)}\sse\cL'$ comprises non-intersecting pairs.
\end{theorem}

\begin{proof}
This follows essentially from Theorem~\ref{combA} and
Lemma~\ref{bapprox}. When $\z_1\leq(1+\e)\cdot\lpopt$, Theorem~\ref{combA}
yields the above bound on $\cost(F_\A)$. Otherwise, if none of the cases in step C3 apply,
then Lemma~\ref{bapprox} bounds $\cost(F_\B)$. In the boundary cases,
when we terminate in step C1 or C3, we have 
$\cost(F)\leq\cost(F_1\sm\f_1)+\cost(\f_1)+12R^*$, which is at most the expression in the
theorem due to part (ii) of Theorem~\ref{thm:outsumr}. 
\end{proof}

\section{Minimizing the maximum radius with lower bounds and outliers} \label{lbksup}

The {\em lower-bounded $k$-supplier with outliers} (\lbkso) problem is the min max-radius
version of \lbksro. The input and the set of feasible solutions are the same as in 
\lbksro: the input is an instance $\I=\bigl(\F,\D,\{L_i\},\{c(i,j)\},k',m\bigr)$, and a
feasible solution is $\bigl(S\sse\F,\sg:\D\mapsto S\cup\{\out\}\bigr)$
with $|S|\leq k$, $|\sg^{-1}(i)|\geq L_i$ for all $i\in S$, and $|\sg^{-1}(\out)|\leq m$.
The cost of $(S,\sg)$ is now 
$\max_{i\in S}\max_{j\in\sg^{-1}(i)}c(i,j)$.
The special case where $m=0$ is called the {\em lower-bounded $k$-supplier} (\lbks)
problem, and the setting 
where $\cD=\cF$ is 
often called the {\em $k$-center} version.

Let $\tau^*$ denote the optimal value; note that there are only polynomially many choices
for $\tau^*$. As is common in the study of $\min$-$\max$ problems, 
we reduce the problem to a ``graphical'' instance, where given some value $\tau$, we try to 
find a solution of cost $O(\tau)$ or deduce that $\tau^*>\tau$. 
We construct a bipartite unweighted graph 
$G_{\tau}=\bigl(V_\tau=\D\cup\F_\tau,E_\tau)$, 
where $\cF_\tau=\{i\in\F:|\cball(i,\tau)|\geq L_i\}$, and $E_\tau=\{ij : c(i,j) \leq \tau , i\in \cF_\tau, j\in \cD\}$. Let
$\dist_{\tau}(i,j)$ denote the shortest-path distance in $G_\tau$ between $i$ and $j$,
so $c(i,j) \leq dist_\tau(i,j) \cdot\tau$. 
We say that an assignment $\sigma:\cD\mapsto\cF_\tau\cup\{\out\}$ is a 
{\em distance-$\alpha$ assignment} if $\dist_\tau(j,\sigma(j)) \leq \alpha$ for every client
$j$ with $\sigma(j)\neq\out$.
We call such an assignment feasible, if it yields a feasible \lbkso-solution, and we say
that $G_\tau$ is feasible if it admits a feasible distance-1 assignment. 
It is not hard to see that given $F\sse\F_\tau$, the problem of finding a feasible
distance-$\al$-assignment $\sg:\cD\mapsto F\cup\{\out\}$ in $G_\tau$ (if one exists) can
be solved by creating a network-flow instance with lower bounds and capacities. 

Observe that an optimal solution yields a feasible distance-1 assignment in $G_{\tau^*}$.
We devise an algorithm that for every $\tau$, either finds a feasible distance-$\al$
assignment in $G_\tau$ for some constant $\al$, or detects that $G_\tau$ is not feasible. 
This immediately yields an $\al$-approximation
algorithm since the smallest $\tau$ for which the algorithm returns a feasible
\lbkso-solution must be at most $\tau^*$. We obtain 
Theorems~\ref{lbksthm} and~\ref{lbksothm} via this template.

\begin{theorem} \label{lbksthm}
There is a $3$-approximation algorithm for \lbks. 
\end{theorem}

\begin{theorem} \label{lbksothm}
There is a $5$-approximation algorithm for \lbkso. 
\end{theorem}

We complement our approximation results via a simple
hardness result (Theorem~\ref{lbkshard}) showing that our approximation factor for \lbks
is tight. We also show that \lbkso is equivalent to the $k$-center version (i.e., where
$\F=\D$) of the problem (Appendix~\ref{append-equiv}); a similar equivalence is known to
hold for the {\em capacitated} versions of $k$-supplier and $k$-center with
outliers~\cite{cygan2014constant}. 

\begin{theorem} \label{lbkshard}
\mbox{It is NP-hard to approximate \lbks within a factor better than $3$, unless $P=NP$.}
\end{theorem}

\begin{proof}
The result is shown via a reduction from set cover problem. Suppose we have a set cover instance with set $\cU = [n]$ of elements and collection $\cS = \cup_{p=1}^{n'}\{S_p\}$ of subsets of $\cU$, and we want to know if there exists $k$ subsets of $\cU$ in $\cS$ that cover all elements of $\cU$. Let $j_1, j_2, \cdots, j_n$ represent the elements and $i_1,i_2,\cdots, i_{n'}$ represent subsets of $\cU$ in $\cS$. Construct an \lbks instance $\cI$ with client set $\cD =\cup_{p=1}^n \{j_p\}$, facility set $\cF = \cup_{q=1}^{n'}\{i_q\}$, define $c(j_p,i_q)$ for $j_{p}\in \cD, i_q\in \cF$ to be $1$ if $p\in S_q$, $3$ otherwise, and let $L_{i} = 1$ for each $i\in \cF$. Suppose  there exists a collection $F$ of $k$ subsets in $\cS$ that cover all elements. First, remove any set $i$ in $F$, if $i$ does not cover an element that is not covered by $F\sm i$. Let $\sigma: \cD \rightarrow F$ be defined for element $j$ to be some set in $F$ that covers $j$. Since each set $i$ in $F$ covers at least one element that is not covered by $F\sm i$, $|\sigma^{-1}(i)| \geq 1$, so $(F, \sigma)$ is a feasible solution to $\cI$ with radius $1$. If no collection of $k$ subsets of $\cU$ in $\cS$ covers all elements, then there does not exist $k$ facilities in $\cF$ that all elements are at distance at most $1$ from them, so optimal solution of $\cI$ has cost at least $3$. Therefore, it is \nphard to approximate \lbks with a factor better than $3$ as otherwise the algorithm can be used to answer the decision problem.
\end{proof}

\paragraph{Finding a distance-$3$ assignment for \lbks.}
Consider the graph $G_{\tau^*}$. Note that there exists an
optimal center among the neighbors of each client in $G$. Moreover, two clients at
distance at least $3$ are served by two distinct centers. 
These insights motivate the following algorithm.

Let $N(v)$ denote the neighbors of vertex $v$ in the given graph $G_\tau$. 
Find a maximal subset $\Gamma$ of clients with distance at least $3$ from each other. If
$|\Gamma|>k$ or there exists a client $j$ with $N(j)=\es$, then return $G_\tau$ is not feasible.
For each 
$j\in\Gamma$, let $i_j$ denote the center in $N(j)$ with minimum lower bound. If there exists a
feasible distance-$3$ assignment $\sigma$ of clients to $F=\bigcup_{j\in \Gamma}\{i_j\}$, return
$\sigma$, otherwise return $G_\tau$ is not feasible. 
The following lemma yields Theorem~\ref{lbksthm}.

\begin{lemma}\label{lem:non-empty-sigma}
The above algorithm finds a feasible distance-3 assignment in $G_\tau$ if $G_\tau$ is feasible.
\end{lemma}

\begin{proof}
Let $\sg^*:\D\mapsto F^*$ be a feasible distance-1 assignment in $G_\tau$.
So $F^*\sse\F_\tau$ and every client has a non-empty neighbor set. 
Since each client in $\Gamma$ has to be served by a distinct center in $F^*$, $|\Gamma|
\leq |F^*| \leq  k$. For each client $j\in \Gamma$, 
let $i^*_j = \sg^*(j)$. Note that $i^*_j\in N(j)$, so $L_{i_j} \leq L_{i^*_j}$ by
the choice of $i_j$, and every client in $\sg^{*-1}(i^*_j)$ is at distance at most $3$
from $i_j$. 

We show that there is a feasible distance-3 assignment $\sg:\D\mapsto F$.
For each $j\in \Gamma$, we assign all 
clients in $\sg^{*-1}(i^*_{j})$ to $i_{j}$. As argued above this satisfies the lower bound
of $i_j$. For any unassigned client $j$, 
let $j'\in \Gamma$ be a client at distance at most $2$ from $j$ (which must exist by
maximality of $\Gamma$). We assign $j$ to $i_{j'}$.
\end{proof}

\paragraph{Finding a distance-$5$ assignment for \lbkso.}
The main idea here is to find a set $F \sse \cF_\tau$ of
 at most $k$ centers that are close to the centers in $F^*\sse\cF_\tau$ for some feasible
 distance-1 assignment $\sg^*:\D\mapsto F^*\cup\{\out\}$ in $G_\tau$.
The non-outlier clients of $(F^*,\sigma^*)$ are close to
 $F$, so there are at least $|\cD|-m$ clients close to $F$. If centers in $F$
 do not share a neighbor in $G_\tau$, then clients in $N(i)$ can be assigned to $i$
 for each $i\in F$ to satisfy the lower bounds. 
We cannot check if $F$ satisfies the above properties, but using an idea similar to
that in~\cite{cygan2014constant}, we will find a sequence of facility sets such that at least one of
these sets will have the desired properties when $G_\tau$ is feasible. 
 
\begin{definition}
Given the bipartite graph $G_\tau$, 
a set $F\subseteq \cF$ is called a {\em skeleton} if it satisfies the following properties.
\begin{enumerate}[(a), itemsep=0.5ex, topsep=0.5ex]
\item ({\em Separation property}) For $i,i'\in F$, $i\neq i'$, we have $\dist_\tau(i,i')
\geq 6$; 

\item There exists a feasible distance-1 assignment $\sg^*:\D\mapsto F^*\cup\{\out\}$ in
$G_\tau$ such that 
\begin{itemize}[nosep]
\item ({\em Covering property}) For all $i^*\in F^*$, $\dist_\tau(i^*,F) \leq 4$, where
$\dist_\tau(i^*,F) = \min_{i\in F} \dist_\tau(i^*,i)$. 
\item ({\em Injection property}) There exists 
$f:F\mapsto F^*$ such that $\dist_\tau(i,f(i)) \leq 2$ for all $i\in F$.   
\end{itemize}
\end{enumerate}
If $F$ satisfies the separation and injection properties, it is called 
a {\em pre-skeleton}. 
\end{definition}

Note that if $F\sse\F_\tau$ is a skeleton or pre-skeleton, then $G_\tau$ is feasible.
Suppose $F\sse\F_\tau$ is a skeleton and satisfies the properties with respect to a
feasible distance-1 assignment $(F^*,\sigma^*)$. 
The separation property ensures that the neighbor sets of any two locations $i,i'\in F$
are disjoint. The covering property ensures that $F^*$ is at distance at most $4$ from
$F$, 
so there are at least $|\cD|-m$ clients at distance at most $5$ from $F$. 
Finally, the injection and separation properties together 
ensure that $|F|\leq k$ since no two locations in $F$ can be mapped to the same location
in $F^*$. Thus, if $F$ is a skeleton, then we can obtain a feasible distance-5 assignment
$\sg:\D\mapsto F\cup\{\out\}$.

\begin{lemma}\label{lem:skel}
Let $F$ be a pre-skeleton in $G_\tau$. 
Define $U=\{i\in \cF_\tau: \dist_\tau(i, F) \geq 6 \}$ and let 
$i = \arg\max_{i'\in U} |N(i')|$. Then, either $F$ is a skeleton, or $F\cup\{i\}$ is a
pre-skeleton.  
\end{lemma}

\begin{proof}
Suppose $F$ is not a skeleton and $F\cup\{i\}$ is not a pre-skeleton. Let
$\sigma^*:\D\mapsto F^*\cup\{\out\}$ be a feasible distance-1 assignment in $G_\tau$ 
such $F$ satisfies the injection property with respect to $(F^*,\sg^*)$. 
Let $f:F\mapsto F^*$ be the mapping given by the injection property.
Since $F\cup\{i\}$ is not a pre-skeleton and $\dist_\tau(i,F) \geq 6$, 
this implies that $\dist_\tau(i,F^*) > 2$, and hence, $\dist_\tau(i,F^*) \geq 4$ as
$G_\tau$ is bipartite. This means that all clients in $N(i)$ are outliers in
$(F^*,\sigma^*)$. Moreover, since $F$ is not a skeleton, there exists a center 
$i^*\in F^*$ with $\dist_\tau(i^*,F) > 4$, and so 
$\dist(i^*,F) \geq 6$. Therefore, $i^*\in U$. By the choice of $i$, we know that 
$|N(i)| \geq |N(i^*)|$. Now consider $F' = F^*\setminus \{i^*\}\cup \{i\}$, and define
$\sigma':\D\mapsto F'\cup\{\out\}$ as follows: $\sg'(j)=\sigma^*(j)$ for all 
$j\notin N(i)\cup N(i^*)$, $\sg'(j)=i$ for all $j\in N(i)$, and $\sg'(j)=\out$ for all
$j\in N(i^*)$. Note that the $F$ covers as many clients as $F^*$, and so 
$\sg':\D\mapsto F'\cup\{\out\}$ is another feasible distance-1 assignment. 
But this yields a contradiction since $F\cup\{i\}$ now satisfies the injection property
with respect to $(F',\sg')$ as certified by the function $f':F\rightarrow F'$ defined by
$f'(s)= f(s)$ for $s\in F$, $f'(i) = i$.    
\end{proof}

If $G_\tau$ is feasible, then $\es$ is a pre-skeleton. A skeleton can have size at most
$k$. So using Lemma~\ref{lem:skel}, 
we can find a sequence $\F'$ of at most $k+1$ subsets of $\cF_\tau$ by starting with 
$\es$ and repeatedly applying Lemma~\ref{lem:skel} until we either have a set of size
$k$ or the set $U$ in Lemma~\ref{lem:skel} is empty. 
By Lemma~\ref{lem:skel}, if $G_\tau$ is feasible then one of these sets must be a
skeleton.  
So for each $F\in\F'$, 
we check if there exists a feasible distance-$5$ assignment 
$\sigma:\D\mapsto F\cup\{\out\}$, and if so, return $(F,\sigma)$. Otherwise we return that
$G_\tau$ is not feasible.

\section*{Acknowledgment}
Part of this work was carried out while the authors were visiting the Hausdorff Institute
of Mathematics (HIM) in Bonn, Germany, as part of the HIM Trimester Program on
Combinatorial Optimization. We thank the organizers of the trimester program, and HIM for
their support and generous hospitality.

\appendix

\section{Improved Approximation Ratio for \lbksr} \label{append-improved}
We now describe in detail the changes to algorithm $\kbalg$ and its analysis leading to
Theorem~\ref{improvthm}.
First, we set $\delta_z=\frac{\e\lpopt}{3n2^n}$ in the binary-search procedure (step B1);
note that the binary search still takes polynomial time. 
By Lemma~\ref{lem:charikaralpha} (specialized to the non-outlier setting), we have
$\|\al^1-\al^2\|_\infty\leq 2^n\dt_z$, which implies that every 
$(i,r)\in T_{1}\cup T_{2}$ is almost tight with respect to $(\alpha^p,\z_p)$ for
$p=1,2$.  
To obtain the improved guarantee, we construct the mapping $\pi:F_1\mapsto F_2$, and
hence, our stars, based on whether pairs $(i',\rad_1(i'))$ and $(i,\rad_2(i))$ intersect
for $i'\in\proj(F_1)$, $i\in\proj(F_2)$.  
To ensure that every $(i',r')\in F_1$ belongs to some star, we first modify $F_2$ and
$T_{2,I}$ by including non-intersecting pairs from $T_{1,I}$ (which are almost tight in
$(\al^2,\z_2)$).  
We consider pairs in $F_1$ in arbitrary order. 
For each $(i,r)\in F_1$, if $(i,\rad_1(i))$ does not intersect any pair
in $T_{2,I}$, we add $(i,\rad_1(i))$ to $T_{2,I}$, add $(i,r)$ to $F_2$, and set
$\rad_2(i)=\rad_1(i)$. We continue this process until all pairs in $F_1$ are scanned or
$|F_2|=k'$.  

\begin{lemma}
If $|F_2| = k'$ after the above process, then $F_2$ is a feasible \ksr solution with
$cost(F_2)\leq(3+\epsilon)\lpopt$, and $T_{2,I}\sse\cL'$ is a set of non-intersecting
pairs. 
\end{lemma}   

\begin{proof}
All clients in $\cD'$ are covered by balls corresponding to the $F_2$-pairs since this
holds even before any pairs are added to $F_2$. 
It is clear that $T_{2,I}\sse\cL'$ and consists of non-intersecting pairs. 
Using Lemma~\ref{lem:charikaralpha}, 
we have $\sum_{(\hat i,\hat r)\in T_{2,I}}(\hat r+\z_1)\leq\sum_{j\in\cD'}\alpha^1_j+2^n\dt_z|T_{2,I}|$,
so $\sum_{(\hat i,\hat r)\in T_{2,I}}\hat r\leq\bigl(1+\frac{\e}{3}\bigr)\lpopt$.
For every $(i,r)\in F_2$ we have $r\leq 3\rad_2(i)$, so
$\cost(F_2)\leq\sum_{(\hat i,\hat r)\in T_{2,I}}3\hat r\leq (3+\epsilon)\lpopt$. 
\end{proof}

So if $|F_2|=k'$ after the above preprocessing, we simply return $(F_2,\rad_2)$.
Otherwise, we combine solutions $F_1$ and $F_2$ using an LP similar to \eqref{cklp}. We
construct a map $\pi:F_1\rightarrow F_2$ similar to before, but with the small
modification that we set $\pi(i',r') = (i,r)$ only if $(i',\rad_1(i'))$ intersects with
$(i,\rad_2(i))$. Due to our preprocessing, $\pi$ is well-defined. 
As before, let star $\nstar_{i,r} = \pi^{-1}(i,r)$ for each $(i,r)\in F_2$.  

\begin{figure}[h!]
\centering
\label{fig:combstep}
\hspace*{-4ex}
\begin{minipage}[b]{0.5\linewidth }
\includegraphics[width=\textwidth]{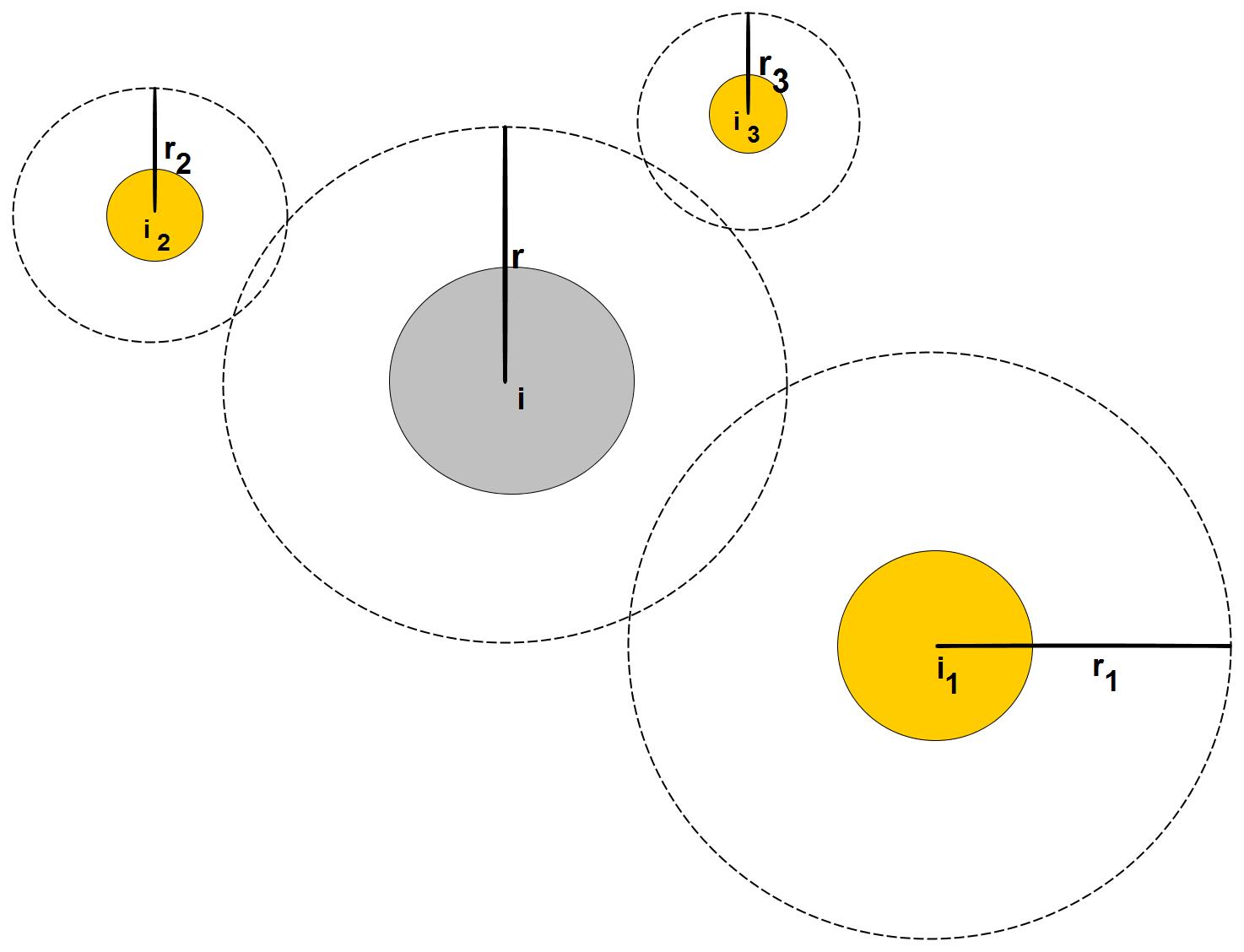}
\caption{Old combination method. 
\mbox{$\nstar_{i,r} = \{(i_1,r_1), (i_2,r_2), (i_3,r_3)\}$}}
\end{minipage}
\quad 
\begin{minipage}[b]{.5\linewidth}
\includegraphics[width=0.95\textwidth]{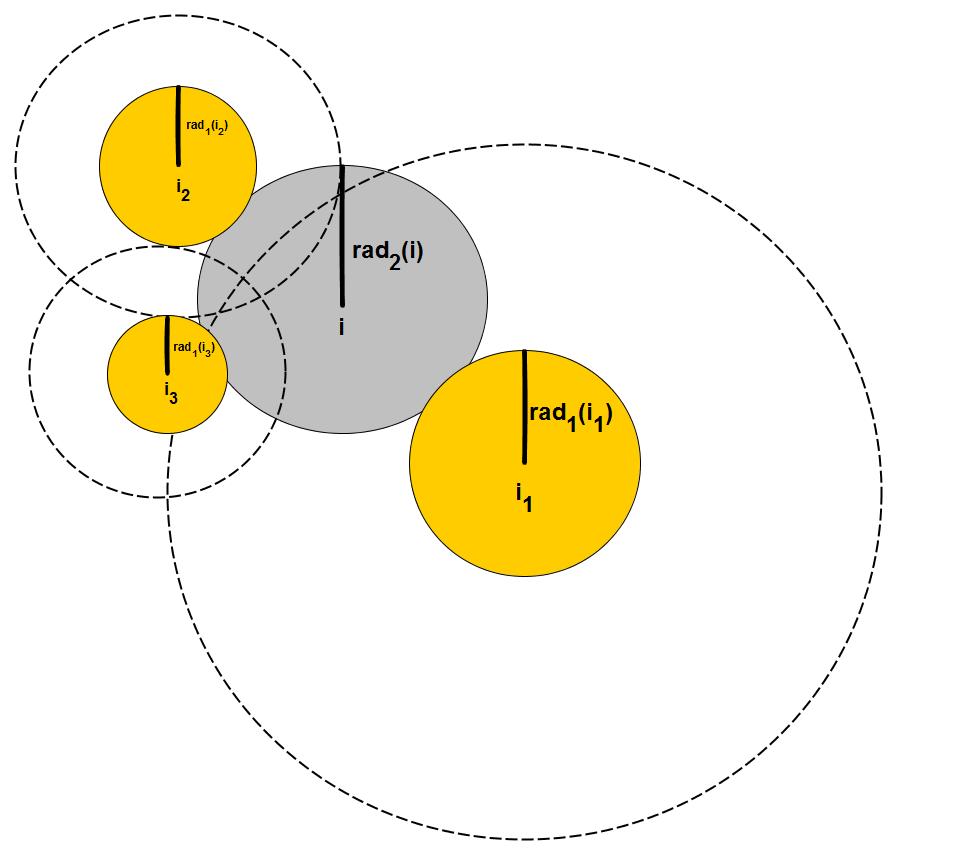}
\caption{New combination method. 
\mbox{$\nstar_{i,r} = \{(i_1,r_1), (i_2,r_2), (i_3,r_3)\}$}}
\end{minipage} 
\end{figure}

The LP again has an indicator variable $x_{i,r}$. If $x_{i,r} = 0$, we select all pairs in
$\nstar_{i,r}$. Otherwise, if $\nstar_{i,r}\neq\es$, we select a pair
$(i',r')\in\nstar_{i,r}$ and include
$\bigl(i',2\rad_2(i)+\sum_{(i'',r'')\in\nstar_{i,r}}4\rad_1(i'')\bigr)$ in our solution;
note that the corresponding ball 
covers all clients in $\bigcup_{(i'',r'')\in\nstar_{i,r}}\cball(i'',r'')$. 
So we consider the following LP.
\begin{alignat*}{2}
\min & \quad & \sum_{(i,r)\in F_2}\Bigl(x_{i,r}\bigl(2\rad_2(i)+
{\textstyle \sum_{(i',r')\in\nstar_{i,r}}}4\rad_1(i')\bigr) & +
(1-x_{i,r}){\textstyle \sum_{(i',r')\in\nstar_{i,r}}} 3\rad_1(i')\Bigr) \tag{C-P'} \label{cklp'} \\
\text{s.t.} & \quad
& \sum_{(i,r)\in F_2}\bigl(x_{i,r}+|\nstar_{i,r}|(1-x_{i,r})\bigr) & \leq k, 
\qquad 0 \leq x_{i,r} \leq 1 \quad \forall (i,r)\in F_2. 
\end{alignat*}

Let $x^*$ be an extreme point of (\ref{cklp'}). Let $F'$ be the pairs obtained by picking
the pairs corresponding to $\ceil{x^*}$ as described above. Since $x^*$ has at most one
fractional component, it follows as before that $|F'|\leq k'$. As before, we return
$\bigl(F',\{\rad(i)\}_{\mu(i)\in F'}\bigr)$ or $(F_2,\{\rad_2(i)\})$, whichever 
has lower cost. 

Let $C'_1 = \sum_{(i,r)\in F_1}\rad_1(i)$ and $C'_2=\sum_{(i',r')\in F_2} \rad_2(i')$. The
following claims are analogous to Claims~\ref{fraccost} and~\ref{clm:costab}.

\begin{claim}\label{clm:fraccost'}
We have $a C'_1 + bC'_2 \leq \bigl(1+\frac{\epsilon}{3}\bigr)\lpopt$.
\end{claim}

\begin{proof}
Using Lemma~\ref{lem:charikaralpha}, we have 
\begin{eqnarray*}
aC'_1+bC'_2 = a \sum_{(i,r)\in F_1} \rad_1(i) + b \sum_{(i,r)\in F_2} \rad_2(i) & \leq &  
a\cdot\Bigl(\sum_{j\in \cD'}\alpha^1_j - k_1\z_1\Bigr) + 
b\cdot\Bigl(k_22^n\delta_z +\sum_{j\in \cD'} \alpha^1_j - k_2\z_2\Bigr) \\
& \leq &\sum_{j\in \cD'} (a\alpha^1_j + b\alpha^1_j) - (ak_1+bk_2)\cdot\z_1 + \frac{\epsilon}{3}\cdot\lpopt \\
& = & \sum_{j\in \cD'} \alpha^1_j-k'\cdot\z_1+\frac{\epsilon}{3}\cdot\lpopt\leq\Bigl(1+\frac{\epsilon}{3}\Bigr)\lpopt.
\qedhere
\end{eqnarray*}
\end{proof}

\begin{claim}\label{clm:costab'}
$\min\{3C'_2, 2bC'_2+(3+b)C'_1\} \leq \frac{3(b+3)}{3b^2-2b+3}(aC'_1+ bC'_2) \leq 3.83(aC'_1+bC'_2) $ for all $a,b \geq 0$ such that $a+b = 1$.
\end{claim}
\begin{proof}
Since the minimum is less than any convex combination,
\begin{eqnarray*}
\min(3C'_2, 2b C'_2+b C'_1 + 3C'_1) &\leq &\frac{3b^2+b}{3b^2-2b+3}(3C'_2) + \frac{-3b+3}{3b^2-2b+3}(2bC'_2+bC'_1+3C'_1) \\
 &= &\frac{3(1-b)(b+3)}{3b^2-2b+3}(C'_1) + \frac{3b(b+3)}{3b^2-2b+3}C'_2 \\
 &= &\frac{3(b+3)}{3b^2-2b+3}((1-b)C'_1+bC'_2).
\end{eqnarray*}

Since $a= 1-b$, the first inequality in the claim follows. 
The expression $\frac{3(b+3)}{3b^2-2b+3}$ is maximized at $b=-3+2\sqrt{3}$, and has value $\frac{3}{8}(5+3\sqrt{3}) \approx 3.8235$, which yields the second inequality in the claim. 
\end{proof}

\begin{lemma} \label{lem:improvcostA} \label{improvcomb}
The cost of the solution $(F,\{\rad(i)\})$ returned by the above combination subroutine is at most
$(3.83+O(\epsilon)) OPT + O(R^*)$ where $\{(i,\rad(i))\}_{i\in \mu(F)} \sse \cL'$ is a set
of non-intersecting pairs. 
\end{lemma}
\begin{proof}
First note that $\{\rad(i)\}$ correspond to $\{\rad_2(i)\}$ if $F= F_2$ and
$\{\rad(i)\}\sse \{\rad_1(i)\}$ if $F=F'$, so in both cases it consists of
non-intersecting pairs from $\cL'$.

The cost of the pair included in $F'$ corresponding to a fractional component of $x^*$ is at most $7R^*$ as each $\rad_p(i)$ is bounded by $R^*$ for $p\in \{1,2\}$. Since $x^*$ has at most one fractional component, $\cost(F')\leq\lpopt_{\text{\ref{cklp'}}}+7R^*$. 
Also, $\lpopt_{\text{\ref{cklp'}}}\leq 2bC'_2+(4b+3a)C'_1=2bC'_2+(3+b)C'_1$, since setting
$x_{i,r}=b$ for all $(i,r)\in F_2$ yields a feasible solution to \eqref{cklp'} of this
cost. 
Therefore, $cost(F)\leq\min\{3C'_2,2bC'_2 + (b+3)C'_1+7R^*\}$, which is at most
$3.83(aC'_1+bC'_2)+7R^*$ by Claim~\ref{clm:costab'}. Combining this with
Claim~\ref{clm:fraccost'} yields the bound in the lemma. 
\end{proof}
 
\begin{proofof}{Theorem~\ref{improvthm}}
It suffices to show that when the selection $F^O=\{(i_1,r_1),\ldots(i_t,r_t)\}$ in step A1
corresponds to the $t$ facilities in an optimal solution with largest radii, we obtain the
desired approximation bound. In this case, if $t=k$, then $F^O$ is an optimal solution;
otherwise, we have $R^*\leq\frac{\iopt}{t}\leq\e\iopt$ and
$\lpopt\leq\iopt-\sum_{p=1}^t r_p$. Combining Lemma~\ref{lem:improvcostA} and 
Lemma~\ref{dcost} then yields the theorem.
\end{proofof}

\section{Proof of Lemma~\ref{lem:charikaralpha}} \label{append-charikar}

We abbreviate $\pdo(\D',\cL',\z)$ to $\pdo(\z)$. We use $x^-$ to denote a quantity
infinitesimally smaller than $x$.
Consider the dual-ascent phase of \pdo for $\z_1$ and $\z_2$. 
First, suppose that $m=0$. Sort clients with respect to their
$\alpha^0_j=\min(\alpha^1_j, \alpha^2_j)$ value. 
Let this ordering be $\alpha^0_1\leq \alpha^0_2\leq \cdots\leq \alpha^0_n$. We prove by
induction that $|\alpha^1_j-\alpha^2_j| \leq 2^{j-1}\delta_z$.  

For the base case, assume without loss of generality that $\alpha^0_j = \alpha^1_j$, and 
let $(i,r)$ be the tight pair that caused $j$ to become inactive in $\pdo(\z_1)$. 
Consider time point $t=\al^0_1$ in the two executions.
By definition all clients are active at time $t^-$ in 
$\pdo(\z_2)$. So the contribution $\sum_{j\in\cball(i,r)\cap\D'}\alpha_j$ of clients
to the LHS of \eqref{open} at time $t^-$ is at least as much as their contribution in
$\pdo(\z_1)$ at time $t^-$. 
Therefore, we can increase $\alpha_1$ by at most $\delta_z$ beyond time $t$ in
$\pdo(\z_2)$ as $z_2-z_1 = \delta_z$. 

Suppose we have shown that for all clients $j=1,2,\cdots,\ell-1$ (where $\ell\geq 2$),
Now consider client $\ell$ and let $(i,r)$ be the tight pair that makes $\ell$ inactive at 
time $\alpha^0_\ell$ in $\pdo(\z_p)$, where $p\in\{1,2\}$. 
Consider time point $t=\al^0_\ell$ in both executions.
By definition, all clients $j>\ell$ are still active at time $t^-$ in both executions
$\pdo(\z_1)$ and $\pdo(\z_2)$. (They might become inactive at time $t$ but can not become
inactive earlier.) 
The contribution $\sum_{j\in\cball(i,r)\cap\D'}\alpha_j$ of clients 
to the LHS of \eqref{open} in the execution other than $p$ at time $t^-$ is at least their
contribution in $\pdo(\z_p)$ at time $t^-$ minus $\sum_{j=1}^{\ell-1} 2^{j-1}\delta_z$. 
The values of $\z$ in the two executions differs by at most $\delta_z$, so in the execution
other than $p$, $\alpha_\ell$ can grow beyond $t$ by at most 
$(1+\sum_{j=1}^{\ell-1} 2^{j-1} )\delta_z \leq 2^\ell\delta_z$. 

Now if we consider a tight pair $(i,r)$ in one of the execution, the value of $RHS$ and
$LHS$ of $\sum_{j\in B(i,r)}\alpha_j \leq r +z$ for the other execution can differ by at
most $(1+\sum_{j=1}^{n} 2^{j-1} )\delta_z \leq 2^n\delta_z$. 

Now consider the case where $m>0$. Note that in this case, we can assume that we have the
execution for $m=0$, pick the first time at which there are at most $m$ active clients, 
i.e., time $\gamma$ in \pdo, and set $\al_j=\gm$ for every active client at this time
point. 
Let $\gamma^0 = \min(\gamma^1, \gamma^2)$, suppose $\gamma^0=\gamma_p$, where 
$p\in\{1,2\}$. Note that by time $\gamma^0+2^n\delta_z$, all pairs that are tight in the
$p$-th execution by time $\gm^0$ are also tight in the other execution. So the number of
active clients after this time point is at most $m$. 
Therefore $|\gamma^1-\gamma^2|\leq 2^n\delta_z$. \qed

\section{\boldmath Equivalence of lower-bounded $k$-supplier with outliers and
  lower-bounded $k$-center with outliers} \label{app:lbkso} \label{append-equiv}
Let \lbkco denote the special case of \lbkso where $\F=\D$. 
In this section, we show that if there exists an $\alpha$-approximation for \lbkco, then
there exists an $\alpha$-approximation for \lbkso. Let $\cI= (k,\cF, \cD, c, L, m)$ be an
instance of \lbkso with $N = |\cF| + 1$ and $|\cD| = n$. Define an instance $\cI'=
(k',\cD\rq{}, c\rq{}, L\rq{}, m\rq{})$ as follows: let $k'=k$ and 
$\cD'=(\cD\times \{1,2,\cdots,N\})\cup \cF$. Let $c'((j,p), i) = c(j,i)$ for each 
$j\in \cD, p\in [N], i\in \cF$, and let $c'$ be the metric completion of these distances 
(i.e., $c'(q,q')$ is the shortest-path distance between $q$ and $q'$ with respect to these
distances for $q,q'\in \cD'$). Define $L'_i = NL_i $ for $i\in \cF$ and 
$L'_{(j,p)} =N(n+1)$, and let $m'=N\cdot m + (N-1)$. Clearly $\cI'$ can be constructed
from $\cI$ in polynomial time. The lower-bounds for $(j,p)$, $j\in \cD, p\in [N]$ are set
so that $L'_{(j,p)}<|\cD'|$, so $(j,p)$ cannot be opened as a center in any feasible
solution to $\cI'$.  

Let $OPT(\cI')$ denote the value of optimal solution of $\cI'$ and $OPT(\cI)$ denote the
value of optimal solution of $\cI$. We claim that $OPT(\cI') \leq OPT(\cI)$. Let
$(F^*,\sigma^*)$ denote an optimal solution of $\cI$. Let solution
$(\hat{F},\hat{\sigma})$ for $\cI$ be constructed as follows: let $\hat{F} = F^*$, for
each $p\in [N]$, define $\sigma(q) = i$ for $q=(j,p)$ if  $\sigma^*(j)=i$, and $\sigma(q)
= \out$ otherwise. Note that since there are at most $m$ outliers in solution
$(F^*,\sigma^*)$ then there are at most $Nm+|\cF|=Nm+ (N-1)$ outliers in
$(\hat{F},\hat{\sigma})$. Clearly the radius of the opened centers is the same as before,
so $OPT(\cI') \leq OPT(\cI)$. 

Now suppose there exists an $\alpha$-approximation algorithm $\cA$ for \lbkco problem. 
Use $\cA$ to generate a solution $(\hat{F},\hat{\sigma})$ for $\cI'$ with maximum radius
$R$. As noted above, we have $\hat{F}\sse \cF$. We construct a solution
$(\hat{F},\sigma')$ for $\cI$ of maximum radius at most $R$ using
Algorithm~\ref{alg:asst}.

\begin{algorithm}
\caption{Constructing a feasible assignment $\sigma'$}
\label{alg:asst}
\begin{algorithmic}[1]
\State Construct network $\cN = (V,E)$ where $V= \{s,t\}\cup \cD \cup \hat{F}$ and $E = \{si: i\in \hat{F}\}\cup \{ij: i\in \hat{F}, j\in \cD, c(i,j)\leq r\} \cup \{jt: j\in \cD\}$.
\State Set $l_{ij}=0$, $u_{ij} = \infty$ for each $ij\in E, i\in \hat{F}, j\in \cD$.
\State Set $l_{si} = L_i$, $u_{si}=\infty$ for each $si \in E, i\in \hat{F}$.
\State Set $l_{jt}=0$, $u_{jt} = 1$ for each $jt\in E, j\in \cD$. 
\State Let $f\gets\text{max-flow}(\cN)$ respecting lower-bounds ($l$) and upper-bounds ($u$) on edges.
\If {value of $f$ is $\geq n-m$}, 
\State set $\sigma'(j) = i $ if $f_{jt} = 1$ and $f_{ij} = 1$ for $i\in \hat{F}$.
\State set $\sigma'(j) = \out$ if $f_{jt} = 0$.
\State \Return $f$.
\EndIf
\State \Return $\sigma' = \emptyset$.
\end{algorithmic}
\end{algorithm}

\begin{lemma}
Solution $(\hat{F}, \sigma')$ is a feasible solution to $\cI$ with maximum radius at most
$R$, where $\sigma'$ is the output of Algorithm $1$. 
\end{lemma}
\begin{proof}
Consider any set $S \sse \hat{F}$. There are at least $\sum_{i\in S} NL_i$ clients in
$\cD'$ assigned to $S$. Since there are at most $N-1$ facilities among $\cD'$, there are
at least $\frac{\sum_{i\in S} N L_i - (N-1)}{N}> \sum_{i\in S} L_i-1$ clients at distance at
most $R$ from $S$. So there are at least $\sum_{i\in S}L_i$ clients in neighbor set of $S$
in $\cN$. It follows that every $s$-$t$ cut in $N$ has capacity at least 
$\sum_{i\in\hat F}L_i$, so there exists a flow $f$ that satisfies the lower-bounds and
upper-bounds on the edges.  

It remains to show that value of $f$ is at least $|\cD|-m$. If there is an incoming edge
to a client in $\cN$, then a flow of $1$ can be sent through $j$. So we want to bound the
number of clients with no incoming edge in $\cN$. If any copy of client $j$ is served
by some facility in the solution $(\hat{F}, \hat{\sigma})$ then $j$ is at distance at most $R$
from some facility in $\hat{F}$. Since there are at most $Nm+(N-1)$ outliers in
$(\hat{F},\hat{\sigma})$, there are at most $\frac{Nm+(N-1)}{N}<m+1$ clients with no incoming
edge in $\cN$. 
\end{proof}

Since algorithm $\cA$ is an $\alpha$-approximation algorithm, wehave 
$R\leq \alpha \cdot OPT(\cI')\leq \alpha OPT(\cI)$. 


\begin{thebibliography}{10}

\bibitem{aggarwal2010achieving}
Gagan Aggarwal, Tom{\'a}s Feder, Krishnaram Kenthapadi, Samir Khuller, Rina
  Panigrahy, Dilys Thomas, and An~Zhu.
\newblock Achieving anonymity via clustering.
\newblock {\em ACM Transactions on Algorithms (TALG)}, 6(3):49, 2010.

\bibitem{AggarwalFKMPZ05}
Gagan Aggarwal, Tom{\'a}s Feder, Krishnaram Kenthapadi, Rajeev Motwani, Rina
  Panigrahy, Dilys Thomas, and An~Zhu.
\newblock Approximation algorithms for k-anonymity.
\newblock {\em Journal of Privacy Technology (JOPT)}, 2005.

\bibitem{AhmadianS16}
Sara Ahmadian and Chaitanya Swamy.
\newblock Approximation guarantees for clustering problems with lower bounds and
  outliers. 
\newblock In {\em Proceedings of the 43rd International Colloquium on Automata, Languages
  and Programming (ICALP)}, 2016.

\bibitem{ahmadian2012improved}
Sara Ahmadian and Chaitanya Swamy.
\newblock Improved approximation guarantees for lower-bounded facility
  location.
\newblock In {\em Proceedings of the 10th International Workshop on
  Approximation and Online Algorithms (WAOA)}, pages 257--271, 2012.

\bibitem{an2015centrality}
Hyung-Chan An, Aditya Bhaskara, Chandra Chekuri, Shalmoli Gupta, Vivek Madan,
  and Ola Svensson.
\newblock Centrality of trees for capacitated k-center.
\newblock {\em Mathematical Programming}, 154(1-2):29--53, 2015.

\bibitem{behsazs15}
Babak Behsaz and Mohammad~R. Salavatipour.
\newblock On minimum sum of radii and diameters clustering.
\newblock {\em Algorithmica}, 73(1):143--165, 2015.

\bibitem{byrka2015improved}
Jaros{\l}aw Byrka, Thomas Pensyl, Bartosz Rybicki, Aravind Srinivasan, and Khoa
  Trinh.
\newblock An improved approximation for k-median, and positive correlation in
  budgeted optimization.
\newblock In {\em Proceedings of the 26th Annual ACM-SIAM Symposium on Discrete
  Algorithms (SODA)}, pages 737--756, 2015.

\bibitem{capoyleasr91}
Vasilis Capoyleas, G{\"u}nter Rote, and Gerhard Woeginger.
\newblock Geometric clusterings.
\newblock {\em Journal of Algorithms}, 12(2):341--356, 1991.

\bibitem{charikar2005improved}
Moses Charikar and Sudipto Guha.
\newblock Improved combinatorial algorithms for facility location problems.
\newblock {\em SIAM Journal on Computing}, 34(4):803--824, 2005.

\bibitem{charikar1999constant}
Moses Charikar, Sudipto Guha, {\'E}va Tardos, and David~B Shmoys.
\newblock A constant-factor approximation algorithm for the k-median problem.
\newblock In {\em Proceedings of the 31st annual ACM Symposium on Theory of
  Computing (STOC)}, pages 1--10, 1999.

\bibitem{CharikarKMN01}
Moses Charikar, Samir Khuller, David Mount, and Giri Narasimhan.
\newblock Algorithms for facility location problems with outliers.
\newblock In {\em Proceedings of the 12th Annual ACM-SIAM Symposium on Discrete
  Algorithms (SODA)}, pages 642--651, 2001.

\bibitem{charikar2001clustering}
Moses Charikar and Rina Panigrahy.
\newblock Clustering to minimize the sum of cluster diameters.
\newblock In {\em Proceedings of the 33rd Annual ACM Symposium on Theory of
  Computing (STOC)}, pages 1--10, 2001.

\bibitem{chen2008constant}
Ke~Chen.
\newblock A constant factor approximation algorithm for k-median clustering
  with outliers.
\newblock In {\em Proceedings of the 19th Annual ACM-SIAM Symposium on Discrete
  Algorithms (SODA)}, pages 826--835, 2008.

\bibitem{cygan2012lp}
Marek Cygan, MohammadTaghi Hajiaghayi, and Samir Khuller.
\newblock Lp rounding for k-centers with non-uniform hard capacities.
\newblock In {\em Proceedings of the 53rd Annual IEEE Symposium on Foundations
  of Computer Science (FOCS)}, pages 273--282, 2012.

\bibitem{cygan2014constant}
Marek Cygan and Tomasz Kociumaka.
\newblock Constant factor approximation for capacitated k-center with outliers.
\newblock In {\em Proceedings of the 31st International Symposium on Theoretical Aspects
  of Computer Science (STACS)}, pages 251--262, 2014.

\bibitem{doddi2000approximation}
Srinivas~R Doddi, Madhav~V Marathe, SS~Ravi, David~Scot Taylor, and Peter
  Widmayer.
\newblock Approximation algorithms for clustering to minimize the sum of
  diameters.
\newblock In {\em Proceedings of the 11th Scandinavian Workshop on Algorithm
  Theory (SWAT)}, pages 237--250, 2000.

\bibitem{ene2013fast}
Alina Ene, Sariel Har-Peled, and Benjamin Raichel.
\newblock Fast clustering with lower bounds: No customer too far, no shop too
  small.
\newblock {\em arXiv preprint arXiv:1304.7318}, 2013.

\bibitem{gibsonk10}
Matt Gibson, Gaurav Kanade, Erik Krohn, Imran~A Pirwani, and Kasturi
  Varadarajan.
\newblock On metric clustering to minimize the sum of radii.
\newblock {\em Algorithmica}, 57(3):484--498, 2010.

\bibitem{gibsonk12}
Matt Gibson, Gaurav Kanade, Erik Krohn, Imran~A Pirwani, and Kasturi
  Varadarajan.
\newblock On clustering to minimize the sum of radii.
\newblock {\em SIAM Journal on Computing}, 41(1):47--60, 2012.

\bibitem{guha2000hierarchical}
Sudipto Guha, Adam Meyerson, and Kamesh Munagala.
\newblock Hierarchical placement and network design problems.
\newblock In {\em Proceedings of 41st Annual IEEE Symposium on Foundations of
  Computer Science (FOCS)}, pages 603--612, 2000.

\bibitem{hochbaums85}
Dorit~S Hochbaum and David~B Shmoys.
\newblock A best possible heuristic for the k-center problem.
\newblock {\em Mathematics of Operations Research}, 10(2):180--184, 1985.

\bibitem{hochbaums86}
Dorit~S Hochbaum and David~B Shmoys.
\newblock A unified approach to approximation algorithms for bottleneck
  problems.
\newblock {\em Journal of the ACM}, 33(3):533--550, 1986.

\bibitem{jain2001approximation}
Kamal Jain and Vijay~V Vazirani.
\newblock Approximation algorithms for metric facility location and k-median
  problems using the primal-dual schema and lagrangian relaxation.
\newblock {\em Journal of the ACM}, 48(2):274--296, 2001.

\bibitem{karger2000building}
David~R. Karger and Maria Minkoff.
\newblock Building steiner trees with incomplete global knowledge.
\newblock In {\em Proceedings of 41st Annual IEEE Symposium on Foundations of
  Computer Science (FOCS)}, pages 613--623, 2000.

\bibitem{khuller2000capacitated}
Samir Khuller and Yoram~J Sussmann.
\newblock The capacitated k-center problem.
\newblock {\em SIAM Journal on Discrete Mathematics}, 13(3):403--418, 2000.

\bibitem{li2013approximating}
Shi Li and Ola Svensson.
\newblock Approximating k-median via pseudo-approximation.
\newblock In {\em Proceedings of the 44th Annual ACM Symposium on Theory of
  Computing (STOC)}, pages 901--910, 2013.

\bibitem{LimWX06}
Andrew Lim, Fan Wang, and Zhou Xu.
\newblock A transportation problem with minimum quantity commitment.
\newblock {\em Transportation Science}, 40(1):117--129, 2006.

\bibitem{mirchandanif90}
Pitu~B Mirchandani and Richard~L Francis.
\newblock {\em Discrete location theory}.
\newblock Wiley-Interscience, 1990.

\bibitem{Samarati00}
Pierangela Samarati.
\newblock Protecting respondents identities in microdata release.
\newblock {\em IEEE Transactions on Knowledge and Data Engineering},
  13(6):1010--1027, 2001.

\bibitem{shmoys}
David~B Shmoys.
\newblock The design and analysis of approximation algorithms.
\newblock {\em Trends in Optimization: American Mathematical Society Short
  Course, January 5-6, 2004, Phoeniz, Arizona}, 61:85, 2004.

\bibitem{svitkina2010lower}
Zoya Svitkina.
\newblock Lower-bounded facility location.
\newblock {\em ACM Transactions on Algorithms (TALG)}, 6(4):69, 2010.

\end{thebibliography}
\end{document}